\documentclass[useAMS,a4,usenatbib]{mn2e}
\input psfig.sty

\usepackage{floatflt, epsfig}
\usepackage{epstopdf}
\usepackage{amsmath}

\def \chisq  {\ifmmode  \chi^2   \else  $\chi^2$  \fi}  
\def \spose#1{\hbox  to 0pt{#1\hss}}  
\def \lta{\mathrel{\spose{\lower 3pt\hbox{$\sim$}}\raise  2.0pt\hbox{$<$}}}
\def \gta{\mathrel{\spose{\lower  3pt\hbox{$\sim$}}\raise 2.0pt\hbox{$>$}}}

\def \kms {\ifmmode  \,\rm km\,s^{-1} \else $\,\rm km\,s^{-1}  $ \fi }
\def \Mpc {\ifmmode  {\rm~Mpc}  \else ${\rm~Mpc}$\fi}  
\def \kpc {\ifmmode  {\rm~kpc}  \else ${\rm~kpc}$\fi}  
\def \pc {\ifmmode  {\rm~pc}  \else ${\rm~pc}$ \fi  }  
\def \Gyr {\ifmmode  {\rm~Gyr}  \else ${\rm~Gyr}$\fi}
\def \Msun {\ifmmode M_{\odot} \else $M_{\odot}$ \fi} 
\def \Lsun {\ifmmode L_{\odot} \else $L_{\odot}$ \fi} 
\def \Rsun {\ifmmode R_{\odot} \else $R_{\odot}$ \fi} 
\def \Msunpyr {\ifmmode M_{\odot}{\rm~yr}^{-1} \else $M_{\odot}{\rm~yr}^{-1}$ \fi} 
\def \hMsun {\ifmmode h^{-1}\,\rm M_{\odot} \else $h^{-1}\,\rm M_{\odot}$ \fi}

\def \LCDM {\ifmmode \Lambda{\rm CDM} \else $\Lambda{\rm CDM}$ \fi}
\def \sig8 {\ifmmode \sigma_8 \else $\sigma_8$ \fi} 
\def \OmegaM {\ifmmode \Omega_{\rm M} \else $\Omega_{\rm M}$ \fi} 
\def \OmegaL {\ifmmode \Omega_{\rm \Lambda} \else $\Omega_{\rm \Lambda}$\fi} 
\def \Deltavir {\ifmmode \Delta_{\rm vir} \else $\Delta_{\rm vir}$ \fi}
\def \rhocrit {\ifmmode \rho_{\rm crit} \else $\rho_{\rm crit}$ \fi}
\def \rhou {\ifmmode \rho_{\rm u} \else $\rho_{\rm u}$ \fi}
\def \zc {\ifmmode z_{\rm c} \else $z_{\rm c}$ \fi}

\def \rhos {\ifmmode \rho_{\rm s} \else $\rho_{\rm s}$ \fi} 
\def \rs {\ifmmode r_{\rm s} \else $r_{\rm s}$ \fi} 
\def \cvir {\ifmmode c_{\rm vir} \else $c_{\rm vir}$ \fi} 
\def \Rvir {\ifmmode r_{\rm vir} \else $R_{\rm vir}$ \fi}
\def \Vvir {\ifmmode V_{\rm  vir} \else  $V_{\rm vir}$  \fi} 
\def \Mvir {\ifmmode M_{\rm  vir} \else $M_{\rm  vir}$ \fi}  
\def \Nvir {\ifmmode N_{\rm  vir} \else $N_{\rm  vir}$ \fi}  
\def \Jvir {\ifmmode J_{\rm vir} \else $J_{\rm vir}$ \fi} 
\def \Evir {\ifmmode E_{\rm vir} \else $E_{\rm vir}$ \fi} 
\def \vvir {\ifmmode v_{\rm vir} \else $v_{\rm vir}$ \fi} 
\def \lam {\ifmmode \lambda  \else $\lambda$ \fi} 
\def \lamp {\ifmmode \lambda^{\prime} \else $\lambda^{\prime}$  \fi} 
\def \Vmax {\ifmmode V_{\rm  max} \else  $V_{\rm max}$  \fi} 
\def \Mdm {\ifmmode M_{\rm  dm} \else $M_{\rm  dm}$ \fi}

\def \Mgas {\ifmmode M_{\rm gas} \else $M_{\rm gas}$ \fi} 
\def \Mcg {\ifmmode M_{\rm cg} \else $M_{\rm cg}$\fi} 
\def \Mhg {\ifmmode M_{\rm hg} \else $M_{\rm hg}$ \fi} 
\def \Mdisc {\ifmmode M_{\rm disc} \else $M_{\rm disc}$ \fi} 
\def \Md {\ifmmode M_{\rm d} \else $M_{\rm d}$ \fi} 
\def \Mda {\ifmmode M_{\rm d,0\%} \else $M_{\rm d,0\%}$ \fi} 
\def \Mdb {\ifmmode M_{\rm d,20\%} \else $M_{\rm d,20\%}$ \fi} 
\def \Mdc {\ifmmode M_{\rm d,40\%} \else $M_{\rm d,40\%}$ \fi} 
\def \md {\ifmmode m_{\rm d} \else $m_{\rm d}$ \fi} 
\def \Mb {\ifmmode M_{\rm b} \else $M_{\rm b}$ \fi} 
\def \Mbh {\ifmmode M_{\rm b,pri} \else $M_{\rm b,pri}$ \fi} 
\def \Mbs {\ifmmode M_{\rm b,sat} \else $M_{\rm b,sat}$ \fi} 
\def \zo {\ifmmode z_{0} \else $z_{0}$ \fi} 
\def \rd {\ifmmode r_{\rm d} \else $r_{\rm d}$ \fi}
\def \rg {\ifmmode r_{\rm g} \else $r_{\rm g}$ \fi}
\def \rb {\ifmmode r_{\rm b} \else $r_{\rm b}$\fi}
\def \rs {\ifmmode r_{\rm s} \else $r_{\rm s}$\fi}
\def \rc {\ifmmode r_{\rm c} \else $r_{\rm c}$\fi}
\def \rvir {\ifmmode r_{\rm vir} \else $r_{\rm vir}$\fi}
\def \rbh {\ifmmode r_{\rm b,pri} \else $r_{\rm b,pri}$ \fi} 
\def \rbs {\ifmmode r_{\rm b,sat} \else $r_{\rm b,sat}$ \fi}

\title[Galactic star formation and accretion histories] 
{Galactic star formation and accretion histories from matching galaxies to dark matter haloes}

\author[B. P. Moster et al.] {Benjamin P. Moster
\thanks{moster@mpa-garching.mpg.de}, Thorsten Naab, Simon D. M. White\\ 
Max-Planck Institut f\"ur Astrophysik, Karl-Schwarzschild Stra\ss e 1, 85748 Garching, Germany\\
}

\begin{document} 
              
\date{\today}
              
\pagerange{\pageref{firstpage}--\pageref{lastpage}}\pubyear{2012} 
 
\maketitle 

\label{firstpage}
             
\begin{abstract}

We present a new statistical method to determine the relationship between the stellar masses
of galaxies and the masses of their host dark matter haloes over the entire cosmic history from
$z\sim4$ to the present. This multi-epoch abundance matching (MEAM) model self-consistently
takes into account that satellite galaxies first become satellites at times earlier than they are observed.
We employ a redshift-dependent parameterization of the stellar-to-halo mass relation to populate haloes
and subhaloes in the Millennium simulations with galaxies, requiring that the observed stellar mass
functions at several redshifts be reproduced simultaneously. We show that physically meaningful
growth of massive galaxies is consistent with these data only if observational mass errors are taken
into account. Using merger trees extracted from the dark matter simulations in combination with MEAM,
we predict the average assembly histories of galaxies, separating into star formation within the galaxies
(in-situ) and accretion of stars (ex-situ). Our main results are: The peak star formation efficiency decreases
with redshift from 23 per cent at $z=0$ to 9 per cent at $z=4$ while the corresponding halo mass
increases from $10^{11.8}\Msun$ to $10^{12.5}\Msun$. The star formation rate of central
galaxies peaks at a redshift which depends on halo mass; for massive haloes this peak is at early cosmic
times while for low-mass galaxies the peak has not been reached yet. In haloes similar to that of the Milky-Way
about half of the central stellar mass is assembled after $z = 0.7$. In low-mass haloes, the accretion of satellites
contributes little to the assembly of their central galaxies, while in massive haloes more than half of the
central stellar mass is formed ex-situ with significant accretion of satellites at $z<2$. We find that our method
implies a cosmic star formation history and an evolution of specific star formation rates which are consistent with
those inferred directly. We present convenient fitting functions for stellar masses, star formation rates, and accretion
rates as functions of halo mass and redshift.

\end{abstract}

\begin{keywords}
cosmology: dark matter, theory --
galaxies: evolution, high-redshift, statistics, stellar content
\end{keywords}

\setcounter{footnote}{1}

\section{Introduction}
\label{sec:intro}

Modern large-scale dark matter simulations are able to make accurate and definite predictions for the
properties of dark matter haloes at all cosmic epochs in cold dark matter (CDM) cosmologies
\citep{springel2005,boylan2008,klypin2011,angulo2012}. In this framework, galaxy formation is driven by the
growth of the underlying large-scale structure and the formation of dark matter haloes: galaxies form by the
cooling and condensation of gas at the centres of the potential wells of such dark matter haloes \citep{white1978,
fall1980}. This results in a tight correlation between the physical properties of galaxies, such as their stellar
mass and star formation rate (SFR), and the mass of the haloes by which they are hosted.

Galaxy formation studies have adopted several approaches to link galaxy properties to those of the dark
matter haloes. The most direct is to model the physical mechanisms that drive the formation
of galaxies from first principles. Here, an $N$-body treatment of the dark matter evolution is combined
with a treatment of the baryonic component based either on a hydrodynamical \citep{katz1996,springel2003,
keres2005,crain2009,schaye2010,oppenheimer2010} or on a semi-analytic method \citep{kauffmann1999,
springel2001,hatton2003,springel2005,kang2005,croton2006,bower2006,guo2011}. Despite substantial
progress, simplified and uncertain recipes must still be employed to model the formation of stars and black
holes and the associated feedback processes, and the detailed properties and evolution of observed galaxy
populations are still only partially reproduced.

An alternative approach to link galaxies to haloes has emerged with the accumulation of data from large galaxy
surveys \citep{cole2001,bell2003,drory2005,fontana2006,panter2007,perez2008,li2009,santini2012}. Instead of
attempting to model the baryonic physics that drive galaxy formation, dark matter haloes are populated with
galaxies using a simple empirical method with adjustable parameters. In the popular halo occupation distribution
(HOD) formalism, the distribution of galaxies (having certain intrinsic properties such as luminosity, color, or type)
within main haloes of a given mass is constrained using galaxy abundance and clustering statistics
\citep[e.g.][]{peacock2000,white2001,berlind2002,zehavi2004,tinker2005,brown2008,zehavi2011,wake2011,leauthaud2012}.
A variant of the classical HOD approach is the conditional luminosity function (CLF) formalism
\citep{yang2003,vdbosch2003,yang2011}. Since reliable galaxy clustering measurements are unavailable at high
redshift, the HOD and CLF approaches have typically been used at low redshift.

In order to circumvent this problem, galaxies can be linked to the underlying substructure directly by assuming a
monotonic relation between stellar mass and halo mass, and matching the cumulative abundance of galaxies to those
of haloes and subhaloes \citep{vale2004,conroy2006,shankar2006,vale2006,moster2010,guo2010,behroozi2010}.
The only observational input to this \lq abundance matching\rq~method is thus the observed stellar mass function (SMF)
or luminosity function. This approach then predicts clustering statistics remarkably well. In this framework, the
stellar-to-halo-mass (SHM) relation can be obtained in two ways: The first \citep[e.g.][]{conroy2006} is to create a mock
galaxy catalogue from the observed SMF for the simulated volume, to order both galaxies and haloes by mass, and then
to match them one by one. A fitting function can then be used to get the SHM relation. The second way
\citep[e.g.][]{moster2010} is first to assume a functional form for the SHM relation, and then to populate the haloes with galaxies and compute
the model SMF. The parameters of the SHM function are then adjusted so that the observed SMF is reproduced. The latter
approach has the advantage that scatter in stellar mass at a given halo mass can easily be added.

The underlying assumptions in this model are that satellite galaxies are living in subhaloes and were central
galaxies before their own haloes were accreted and became subhaloes. The properties of the galaxy population
today are thus a consequence of the formation of central galaxies in main haloes at different redshifts and the
subsequent accretion and evolution of satellites. Since the subhalo is affected much more by tidal stripping than the
galaxy at its centre, the stellar mass of satellites is more closely related to subhalo mass at the time of accretion
rather than to current subhalo mass. Most abundance matching studies simply assign the infall mass to subhaloes.
As pointed out by \citet{yang2011}, this is equivalent to assuming that the SHM relation does not evolve with redshift,
i.e. that the relation between the stellar mass of central galaxies and their halo mass today is the same as the relation
between the stellar mass of satellite galaxies and their halo mass at the time of accretion. However, applying
the technique at different redshifts results in an evolving SHM relation \citep[e.g.][]{conroy2006,moster2010}. This
indicates that the classical abundance matching approach is not self-consistent.

This problem can be solved if we assign the stellar mass of the satellites depending on the infall mass of their subhalo
\textit{and} the infall redshift. The stellar mass of the central galaxies still depends on the main halo mass at the
redshift of observation in the same way. This is equivalent to assuming that the stellar mass of a satellite does not change,
once it has entered the main halo. In order to implement this scheme, we first use the classical abundance matching approach
to see how the parameters of the SHM relation depend on redshift and find a redshift-dependent parameterization of the
SHM relation $m_*(M_{\rm h},z)$. We then employ this parameterization to put a galaxy at the centre of every halo and subhalo
in the simulation at all redshifts between $z=0$ and $z=4$, adjusting the parameters of the relation to reproduce simultaneously
all the observed SMFs in this redshift range. In this way we use a self-consistent relation between stellar mass, halo mass and
redshift for both centrals and satellites.

Statistical models, such as the abundance matching approach, are typically employed to derive constraints on the
relation between galaxy properties and halo mass at a given epoch. With the advent of wide high-redshift surveys,
it has become possible to use this connection, together with information on the evolution of dark matter haloes, to study
the evolution of galaxy properties throughout their history \citep[e.g.][]{white2007,conroy2007}. In particular, \citet{conroy2009}
show how this method can be extended to put constraints on the SFR of galaxies. For this they use an analytic
prescription for galaxy and halo abundances and the evolution of haloes, focusing on $z\lta1$. We follow a similar
approach, but use $N$-body simulations to get halo abundances and assembly histories. Then
we identify the progenitors of a given halo at an earlier epoch and derive the amount of stellar mass that has been
accreted onto the central galaxy and the total stellar mass by which it grew. Finally, employing a recipe to account
for stellar mass loss, we compute the stellar mass that has been formed through star formation and convert this
into a SFR. We are thus able to derive the star formation history (SFH) of the galaxy at the centre of the halo.

This paper is organized as follows. In Section 2, we describe the $N$-body simulations including the halo finding
algorithm and the treatment of \lq orphan\rq~ galaxies. We also summarize the observational SMFs that
have been used. In Section 3 we describe the details of the model that matches galaxies to haloes. In Section 4 we use this relation and the
halo accretion histories to derive SFR histories as a function of halo mass and we show that our model can reproduce the
observed evolution of the cosmic SFR density and of the specific SFRs of galaxies. We discuss some
of the implications of our model in Section 5 and summarize in Section 6.

Throughout this paper, we assume a \lq\textit{WMAP}7\rq~$\Lambda$CDM cosmology with $(\Omega_{\rm m},\Omega_{\rm \Lambda},
\Omega_{\rm b},h,n,\sigma_8)= (0.272, 0.728, 0.046,0.704,0.967,0.810)$. We employ a \citet{chabrier2003} initial
mass function (IMF) and we convert all stellar masses to this IMF. All virial masses are computed with respect to 200
times the critical density. In order to simplify the notation, we will use the capital $M$ to denote dark matter halo
masses and the lower case $m$ to denote galaxy stellar masses.

\section{Simulations and Observations} 

In this paper we aim to link observable properties of galaxies to the underlying dark matter distribution
using a statistical sample of both galaxies and dark matter haloes.
The properties of the halo population originate from two $N$-body simulations: the Millennium
Simulation \citep[MS][]{springel2005} for massive haloes and the Millennium-II Simulation
\citep[MS-II][]{boylan2009} for lower mass haloes.

For the galaxy population at low redshift we use the precise stellar mass functions of \citet{li2009} based on the
spectroscopic data set of the Sloan Digital Sky Survey \citep[SDSS][]{york2000}. At high redshift
we use two sets of observed SMFs: for massive galaxies we use the SMFs by
\citet[][PG08]{perez2008} taken from a wide survey, and for low-mass galaxies we use the SMFs by
\citet[][S12]{santini2012} taken from a deep survey.

\subsection{Numerical Simulations} 
\label{sec:nsim}

Both Millennium simulations adopt the concordance $\Lambda$CDM cosmology with parameters chosen to agree
with the first-year \textit{Wilkinson Microwave Anisotropy Probe} (\textit{WMAP}) data \citep{spergel2003}. As these
parameters are not consistent with the latest \textit{WMAP} data analysis, we use the scaling recipe by \citet{angulo2010}
to convert the properties of the simulations to represent the growth of structure in a \textit{WMAP}7 cosmology.
Specifically, we chose $\Omega_{\rm m}=0.272$, $\Omega_{\rm \Lambda}=0.728$, $\Omega_{\rm b}=0.0455$,
$H_{\rm 0}=70.4{\rm~km}{\rm~s}^{-1}{\rm~Mpc}^{-1}$, $n=0.967$ and $\sigma_8=0.81$ \citep{komatsu2011}.

Both simulations were carried out in a periodic box advancing $2160^3$ dark matter particles from redshift $z\sim127$ to
0. The side length of the box is $761\Mpc$ for the \textit{MS}, and $152\Mpc$ for the \textit{MS-II}. The corresponding
particle masses are $1.58\times10^9\Msun$ and $1.27\times10^7\Msun$, respectively. The large volume of the 
\textit{MS} enables us to sample reliably even the central galaxies of rare and massive clusters, while the \textit{MS-II} has
sufficient resolution to sample the faintest dwarf galaxies.

Dark matter haloes are identified at each output using a friends-of-friends (FoF) halo finder that links together
particles with separation less than 0.2 of the mean interparticle distance. Substructures inside the FoF groups are
then identified using {\sc subfind} \citep{springel2001}. The most massive self-bound subgroup is
defined as the \lq main halo\rq, and all other subgroups within the FoF groups are referred to as \lq subhaloes\rq.
The virial radius and mass of the main haloes are determined with a spherical overdensity criterion: the density
inside a sphere centered on the most bound particle is required to be equal to 200 times the critical value.
Merger trees were then built to link each subhalo found at a given snapshot (including the main haloes) to a unique
descendant in the following snapshot. With these merger trees we can track the evolution of each halo throughout
the simulation.

In this work, we assume that both main haloes and subhaloes host galaxies at their centres. We will refer to the
galaxy at the center of a distinct halo as a \lq central galaxy\rq, and the galaxies within subhaloes as \lq satellites\rq.
There is thus one central galaxy within each FoF group but there may be any number of sattelites (including zero).
Ab initio models of galaxy formation predict that the stellar mass of a galaxy is tightly correlated with the depth of the
potential well of the halo in which it formed. For central galaxies, the relevant mass is the virial mass of the main halo
at the time of observation, but for satellites the situation is more complex: Subhaloes lose mass while orbiting in a larger
system as their outer regions are tidally stripped. Stars are centrally concentrated and more tightly bound than the dark
matter, however, so the stellar mass of a satellite galaxy changes only slightly after it enters a larger halo. Therefore
the subhalo mass at the time of observation is not a good tracer for the stellar mass of a satellite. A better tracer is the subhalo
mass at the time that it was accreted by the main halo, which typically corresponds to the maximum mass over its history.

Due to the finite mass resolution of the simulations, subhaloes can no longer be identified once their mass has been tidally
stripped below the resolution limit. Since mass loss can be substantial this is important even for fairly massive
subhaloes. A special treatment of these so-called \lq orphans\rq~ is necessary. We determine the orbital parameters the
last time a subhalo is identified in the simulation and use them in the dynamical friction formula of \citet{binney1987}:
\begin{equation}
t_{\rm df}=\alpha_{\rm df}\frac{V_{\rm vir}r^2_{\rm sat}}{G M_{\rm sat} \ln \Lambda}\;,
\end{equation}
where $r_{\rm sat}$ is the distance between the centers of the main halo and of the subhalo at the time the merger clock was set,
$M_{\rm sat}$ is the mass of the subhalo at this time, and $\ln\Lambda=\ln(1+\Mvir/M_{\rm sat})$ is the Coulomb
logarithm. Following \citet{delucia2007} we set $\alpha_{\rm df}=2.34$, which was also found to be appropriate in the
$N$-body studies by \citet{boylan2008}. We retain the disrupted subhalo for the purposes of abundance matching until a time
$t_{\rm df}$ has elapsed, and assume that only then does it truly merge with the main halo.

\subsection{Observed stellar mass functions}

Having derived the abundance of the halo/subhalo population as a function of mass and redshift, we now have to select samples
of observed galaxy abundances, to link the two populations. At low redshift ($z=0.1$), the SDSS galaxy catalogues allow a determination
of the local SMF down to a stellar mass of $10^7\Msun$. Above about $10^8\Msun$ it is robust against incompleteness.
The large volume of the sample leads to very low uncertainties due to cosmic variance \citep{trenti2008,moster2011}.

We employ the recent measurements by \citet{li2009}, based on a complete and uniform sample from the SDSS/DR7
\citep{abazajian2009}, and we use the conversion to stellar mass based on the SDSS $r$-band model luminosities, as
derived by \citet{guo2010}. This SMF ranges from stellar masses of $10^{8.3}\Msun$ to $10^{11.7}\Msun$. In order to
extend the SMF to even lower stellar masses, we include the measurements by \citet{baldry2008} between $10^{7.4}\Msun$
and $10^{8.3}\Msun$.

As we aim at constraining the stellar mass content of haloes back to $z=4$ we require high-quality observed SMFs at high redshift. In general
wide high-redshift surveys capture many massive galaxies but are not deep enough to constrain the low-mass tail of the SMF. Conversely,
deep surveys cover too small volumes to constrain the massive tail. In practice obtaining a SMF that is equally well constrained for massive and
low-mass galaxies is difficult, as this would involve a survey that is both wide (capturing many massive galaxies) and deep (detecting even
galaxies of very low mass).

We use the mass functions of the wide survey of \citet{perez2008} to constrain the massive end. These SMFs have been derived in 12 redshift bins
from a sample of galaxies at $0<z<4$ selected in three different fields at $3.6-4.5~\mu m$
with the Infrared Array Camera \citep{fazio2004} on board the Spitzer Space Telescope \citep{werner2004}. The survey
covers a total area of $664{\rm~arcmin}^2$ and the obtained galaxy catalogue is 90 per cent complete down to $\sim5~\mu{\rm Jy}$ at
$3.6~\mu m$ and $\sim4~\mu{\rm Jy}$ at $4.5~\mu m$. Due to the wide area of the survey, the SMFs put tight constraints on the
abundances of massive galaxies up to $z=4$, allowing for detailed modeling of the evolution of such galaxies.

For the low-mass end we use SMFs in 6 redshift bins by \citet{santini2012}, which have been obtained from Early Release
Science observations taken in the $Y$, $J$, and $H$-bands with the near-IR Wide Field Camera 3 on board HST \citep{windhorst2011},
and from deep $K$-band images taken with the near-IR VLT imager Hawk-I \citep{castellano2010}. The galaxy catalogue has been
selected in the $K$-band, covers an area of $33{\rm~arcmin}^2$, and is complete to a typical magnitude of $K=25.5$. This very deep,
albeit narrow, survey is able to resolve galaxies of low mass up to $z=4.5$, and allows us to study the evolution of low-mass galaxies
in detail. For all SMFs we have converted the stellar masses to agree with a Chabrier IMF. An overview of the 19 SMFs used in this work
and their redshifts is shown in Figure \ref{fig:smfS}.

\section{Connecting galaxies and haloes} 
\label{sec:method}

\begin{figure}
\psfig{figure=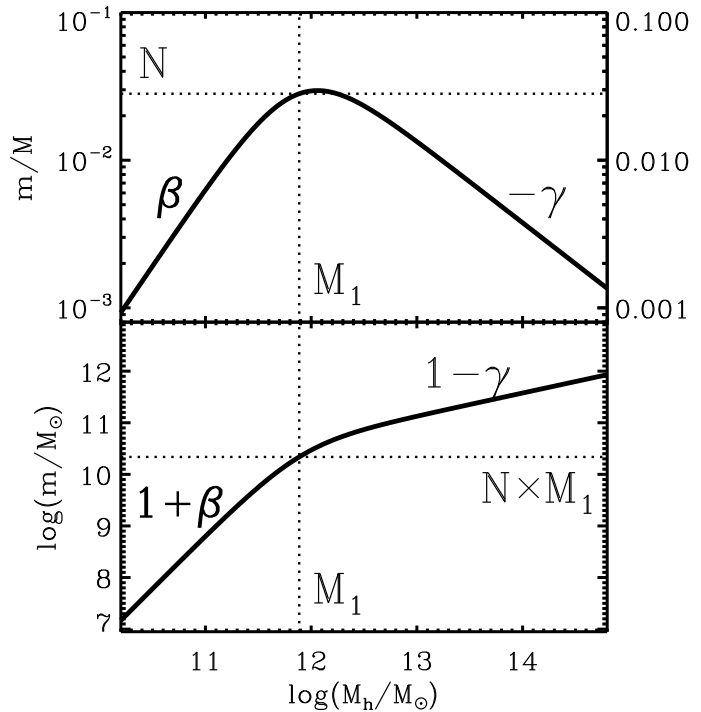,width=0.45\textwidth}
\caption{\textit{Upper panel}: Sketch of the stellar-to-halo mass ratio as a function of halo mass peaking around the
characteristic mass $M_1$ where it has the normalization $N$. It has a low-mass slope $\beta$ and a high mass slope
$-\gamma$. \textit{Lower panel}: Sketch of the stellar-to-halo mass relation as a function of halo mass. The low-mass slope
is $1+\beta$ and the high mass slope is $1-\gamma$.}
\label{fig:sketch}
\end{figure}

In this section we describe how we derive the relationship between the stellar mass of a galaxy
and the mass of its dark matter halo. For this we first need to specify the
SHM ratio. A direct comparison of the halo mass function and the stellar mass function shows that the
baryons are converted into stars with very different efficiencies in haloes of different mass. Star formation
is most efficient in systems like the Milky Way with $\Mvir\sim10^{12}\Msun$, and is substantially less efficient
in much more massive or
much less massive haloes. We adopt the parameterization of the SHM ratio of \citet{moster2010}:
\begin{equation}
\label{eqn:shmratio}
\frac{m}{M} = 2 \; N \; \left[ \left( \frac{M}{M_1} \right)^{-\beta} + \left( \frac{M}{M_1} \right)^{\gamma} \right]^{-1} .
\end{equation}
It has four free parameters: the normalization of the SHM ratio $N$, a characteristic mass $M_1$, where the ratio is equal
to the normalization $N$, and two slopes $\beta$ and $\gamma$ which indicate the behavior of $m/M$ at the low and high-mass
ends, respectively. Figure \ref{fig:sketch} illustrates the SHM relation in two forms, indicating the four parameters.
The halo mass where the SHM ratio is maximal is related to the characteristic mass $M_1$ by
\begin{equation}
\label{eqn:maxratio}
M_{\rm max} = M_1 \left(\frac{\beta}{\gamma}\right)^{1/(\beta+\gamma)}\,.
\end{equation}

This SHM ratio can then be used to compute the average stellar mass in a dark matter halo. However, we
expect that two haloes of the same mass $M$ may harbor galaxies with different stellar masses, since they
can have different halo concentrations, spin parameters, and merger histories. Therefore we assume an intrinsic
scatter around this average relation when assigning the stellar mass to an individual galaxy: we draw a stellar mass $m$
from a lognormal distribution with a mean value given by equation \eqref{eqn:shmratio} and with a
scatter of $\sigma_m(\log m)=0.15$.

\subsection{The evolution of the stellar-to-halo mass relation} 
\label{sec:matching}

\begin{figure*}
\psfig{figure=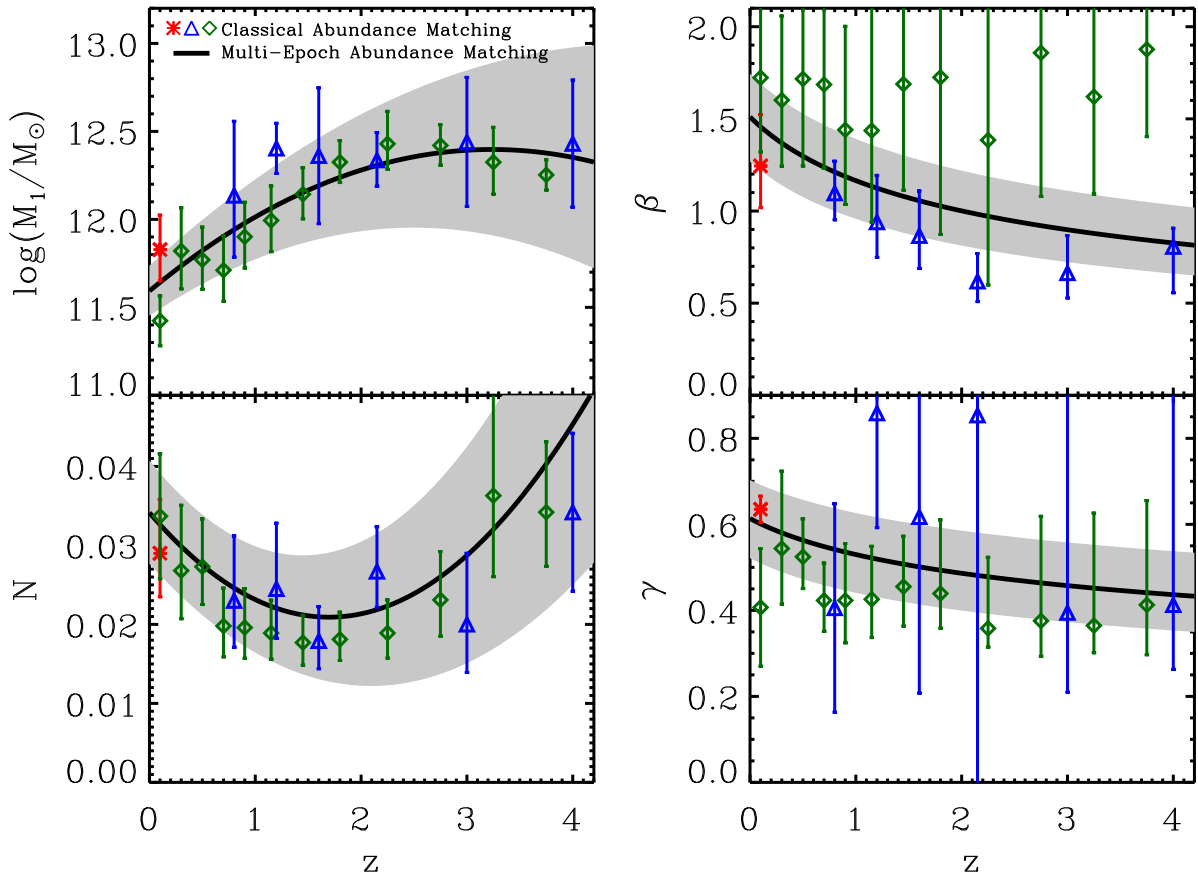,width=0.9\textwidth}
\caption{Evolution of the SHM relation parameters with redshift in a model without observational mass errors. The symbols
correspond to the values that have been derived with the classical abundance matching approach at individual redshifts.
Different colors represent the different SMFs that have been used to derive the SHM relation: red crosses for the SDSS SMF,
green diamonds for the PG08 SMFs and blue triangles for the S12 SMFs. The solid line corresponds to a multi-epoch abundance
matching model that takes into account that satellites are accreted at different epochs. The shaded area indicates the $1\sigma$
confidence levels. For $M_1$ and $N$ we assume a second order polynomial in $z$ and for $\beta$ and $\gamma$ a power
law in $z$.}
\label{fig:parameters}
\end{figure*}

As a first step, we investigate how the parameters of the SHM relation evolve with redshift. For this we
assume that at a given redshift the relation between the stellar mass of a satellite galaxy and the
maximum mass of its dark matter halo over its history is the same as the SHM relation of central galaxies.
This assumption is only an approximation as the stellar mass of satellites is related to the halo mass at infall
and the SHM relation is expected to have changed since this epoch. However, at any redshift the number of subhaloes
of a given mass is lower than the number of main haloes of the same mass, so the approximation is not greatly in error.
With this approach we are able study how the parameters of the SHM relation depend on redshift, and which functional
form may be appropriate.

In order to constrain the parameters at a given redshift (where an observed SMF is available), we populate the
haloes in the simulation with galaxies using the SHM relation to derive their stellar masses. Once the simulation
box is filled with galaxies, it is straightforward to compute the model SMF. We then assign a likelihood to this
model SMF by relating it to the observed SMF at this redshift:
\begin{eqnarray}
\mathcal{L} &=& \exp\left(-\chi^2_r\right) \nonumber\\
\chi^2_r&=& \frac{1}{N_\Phi} \sum_{i=1}^{N_\Phi} \left( \frac{\log\Phi_{\rm mod}(m_i)-\log\Phi_{\rm obs}(m_i)}{\sigma_{\rm obs}(m_i)} \right)^2 \,.
\end{eqnarray}
Employing a Markov chain Monte Carlo (MCMC) method, we sample the probability distribution for the parameters
and extract the best-fit values and their $1\sigma$ errors. We repeat this procedure for every observed SMF
available and plot the results in Figure \ref{fig:parameters} (symbols with errorbars).

The transition mass $M_1$ increases with increasing redshift and reaches a maximum around $z\sim2$, after which
it stays constant or even slightly decreases again. This means that at $z=0$ the conversion of baryons into stars has
been most efficient in haloes with a virial mass of $\log(\Mvir/\Msun)\sim11.7$, while at higher redshift the conversion
was most efficient in more massive haloes with $\log(\Mvir/\Msun)\sim12.3$. The normalization $N$, i.e. the SHM ratio at the 
most efficient halo mass, decreases with increasing redshift, reaches a minimum around $z\sim2$ and then increases
again. We note that this does not imply that individual galaxies lose mass as halo mass generally increases for an individual
system.

The low-mass slope $\beta$ is only constrained by the S12 SMFs. This is a consequence of the properties
of the surveys from which the SMFs are obtained: The survey in PG08 does not provide strong constraints on the low-mass
slope of the SMFs and consequently on $\beta$. The survey used in S12, on the other hand, is able to detect
low-mass galaxies. From these constraints, we see that $\beta$ is decreasing with redshift, which means that the slope of
the SHM relation becomes shallower with increasing redshift. For the high mass slope $\gamma$ constraints are
provided only by the PG08 SMFs. These constraints indicate that $\gamma$ is also decreasing with redshift, i.e.
the slope of the SHM relation becomes steeper with increasing redshift.

\subsection{The multi-epoch abundance matching model}
\label{sec:asm}

\begin{figure}
\psfig{figure=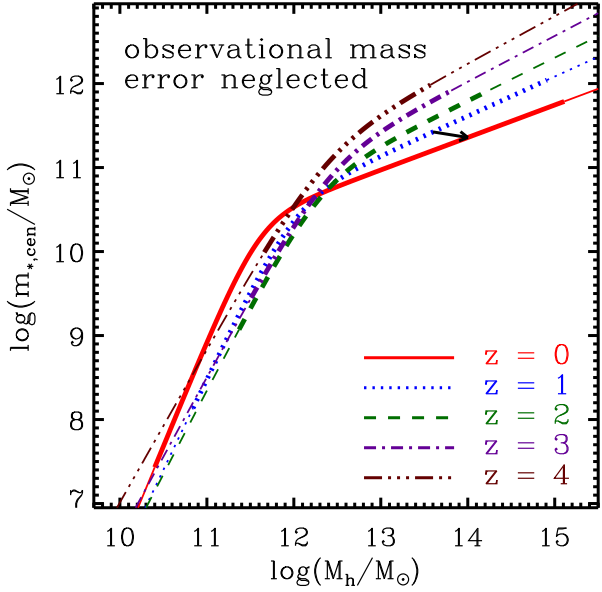,width=0.45\textwidth}
\caption{Stellar mass of the central galaxy as a function of the virial mass of the main halo for different redshifts for the model
without observational mass errors. The thick lines correspond to the ranges where the relation is constrained by data, while the
thin lines are extrapolations. The arrow shows the evolution of a system with $\Mvir(z=0)=10^{14}\Msun$ from $z=1$
to 0, indicating that if observational mass errors are not accounted for, massive systems lose significantly more stellar mass than
expected from stellar evolution models.}
\label{fig:sheff}
\end{figure}

As we now have a first idea how the SHM parameters should evolve with redshift, we can construct a self-consistent model. As the next
step in this multi-epoch abundance matching (MEAM) model, we determine a convenient redshift parameterization of the SHM parameters. 
For the characteristic mass $M_1$ and the normalization
$N$, we use a second order polynomial in $z$, while for the slopes $\beta$ and $\gamma$ we use a power law:
\begin{eqnarray}
\log M_1(z) &=& M_{10}+M_{11}z+M_{12}z^2,\\
N(z) &=& N_{10}+N_{11}z+N_{12}z^2,\\
\beta(z) &=& \beta_{10}(z+1)^{\beta_{11}},\\
\gamma(z) &=& \gamma_{10}(z+1)^{\gamma_{11}}.
\end{eqnarray}

The stellar mass of a central galaxy is then given as a function of the virial mass of its dark matter halo \Mvir and redshift $z$.
For satellite galaxies, the stellar mass is now determined by employing the SHM relation at the time at which it still was a
central galaxy, i.e. at the redshift of infall $z_{\rm inf}$. We then assume that the stellar mass of a satellite does not change after its subhalo
entered the main halo, i.e. we neglect stellar stripping and star formation in the satellite. We thus get a redshift-dependent SHM relation
$m(M,z)$ that depends on the virial mass of the main halo and redshift at the current epoch for central galaxies and on the infall mass
of the subhalo and infall redshift for satellite galaxies. The stellar mass of a galaxy at redshift $z$ is then given by

\begin{equation}
\label{eqn:censat}
m = \begin{cases}
          m(M_{\rm vir},z) & \text{for host haloes}\\
          m(M_{\rm inf},z_{\rm inf}) & \text{for subhaloes}\;.
       \end{cases}
\end{equation} 
In this way we can derive the stellar mass of any galaxy just by employing the redshift-dependent SHM relation for central galaxies.

This relation can then be constrained with a set of SMFs at different redshifts. To this end we employ the simulation at all redshifts where
an observed SMF is available and populate all main haloes and subhaloes in the simulation with galaxies using the redshift-dependent
SHM relation to derive their stellar masses. We then compute the $N_z$ model SMFs for each redshift and assign a likelihood to the model
by relating them to the corresponding observed SMFs:
\begin{eqnarray}
\mathcal{L} &=& \exp\left(-\chi_{\rm r}^2\right) \nonumber\\
\chi_{\rm r}^2 &=& \frac{1}{N_z} \sum_{i=1}^{N_z} \chi_{z,i}^2 \nonumber\\
\chi_{z,i}^2 &=& \frac{1}{N_{\Phi,i}} \sum_{j=1}^{N_{\Phi,i}} \left( \frac{\log\Phi_{\rm mod}(m_j)-\log\Phi_{\rm obs}(m_j)}{\sigma_{\rm obs}(m_j)} \right)^2.
\end{eqnarray}
We again employ an MCMC method, sample the probability distribution for all the parameters and extract the best-fit values and their $1\sigma$ errors.
We show the results in Figure \ref{fig:parameters} for the simultaneous fit to all SMFs (black line and shaded area). For all four SHM parameters, the lines
for the simultaneous fit follow the trends that have been seen for the fits at individual redshifts. 

The SHM relation for a set of redshifts between $z=0$ and $z=4$ is plotted in Figure \ref{fig:sheff}. At a fixed
halo mass at the high-mass end ($\log(\Mvir/\Msun)\gta13.0$), the stellar mass increases with with
increasing redshift. At intermediate halo masses ($\log(\Mvir/\Msun)\sim12.0$) the stellar mass decreases with increasing redshift, reaches a minimum
at $z\sim2$ and then increases again (reflecting the evolution of the normalization $N$). The fact that at a given halo mass the mass of the central
galaxy is lower at present than at an earlier epoch does not necessarily imply that individual galaxies lose mass during their evolution, as haloes also
become more massive over time. At a fixed halo mass of $\log(\Mvir/\Msun)=12.0$ the model stellar mass is $\log(m_*/\Msun)=10.5$ at $z=4$ and only
$\log(m_*/\Msun)=10.2$ at $z=2$. However, a typical halo with $\log(\Mvir/\Msun)=12.0$ at $z=4$ will increase its virial mass up to
$\log(\Mvir/\Msun)=12.7$ at $z=2$, for which the model stellar mass is $\log(m_*/\Msun)=11.0$, so that the stellar mass of the central galaxy does not
decrease, but increase.

\begin{figure*}
\psfig{figure=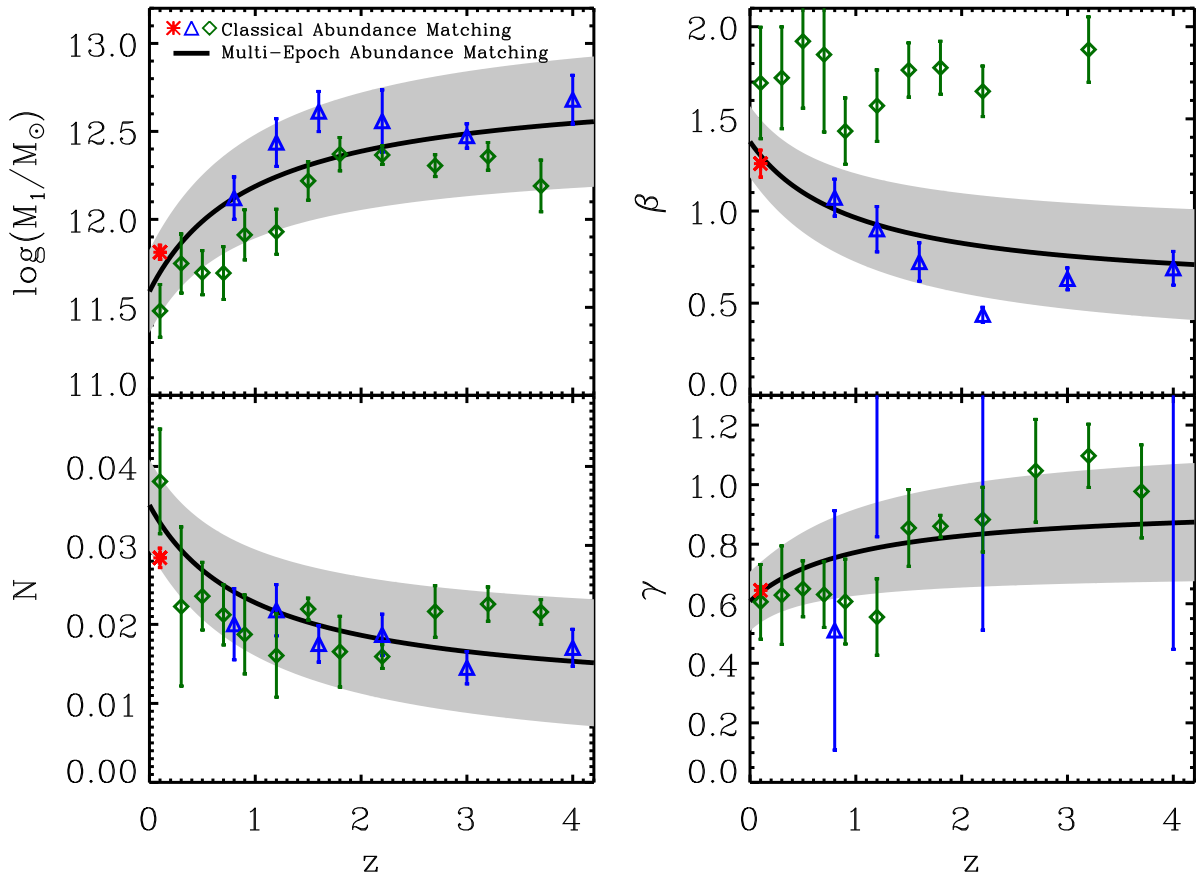,width=0.9\textwidth}
\caption{Evolution of the SHM relation parameters with redshift for the model including observational mass errors. The symbols of different color
correspond to the classical abundance matching approach at individual redshifts: red crosses for the SDSS SMF, green diamonds for the PG08
SMFs and blue triangles for the S12 SMFs. The solid line corresponds to the multi-epoch abundance matching model. The shaded area indicates the $1\sigma$
confidence levels. For all four parameters we assume a linear dependence on the expansion factor $a=(1+z)^{-1}$.}
\label{fig:parametersS}
\end{figure*}

However, we encounter a severe problem for massive galaxies at low redshift: the growth of halo mass cannot compensate for the decreasing
stellar mass. A typical massive halo of $\log(\Mvir/\Msun)=14.0$ at $z=0$ has grown from a $z=1$ halo with a virial mass of $\log(\Mvir/\Msun)=13.6$.
The model stellar mass for the central galaxy of this system at $z=1$ is $\log(m_*/\Msun)=11.4$, while at $z=0$ it is only $\log(m_*/\Msun)=11.3$. This would
imply that massive systems lose stellar mass at low redshift and is unphysical, as stellar mass loss rates from supernovae are much lower than accretion
rates onto the central galaxy. This indicates that an important effect has been neglected in our model and that the effects of this missing ingredient
are most severe at the massive end. In the following section we show that neglecting errors in the observed stellar mass impacts the results
for massive galaxies strongly and that taking this into account results in a model where galaxies do not lose stellar mass during their evolution.

\subsection{Introducing observational mass errors in the model}
\label{sec:ome}

\begin{table*}
\begin{center}
 \begin{minipage}{100mm}
  \caption{Fitting results for the redshift-dependent stellar-to-halo mass relationship.}
  \begin{tabular}{@{}lcccccccc@{}}
  \hline
   & $M_{10}$ & $M_{11}$ & $N_{10}$ & $N_{11}$ & $\beta_{10}$ & $\beta_{11}$ & $\gamma_{10}$ & $\gamma_{11}$\\
 \hline
Best fit & 11.590 & 1.195 & 0.0351 & -0.0247 & 1.376 & -0.826 & 0.608 & 0.329\\
$\sigma$ & ~0.236 & 0.353 & 0.0058 & ~0.0069 & 0.153 & ~0.225 & 0.059 & 0.173\\
\hline
\label{t:pars}
\end{tabular}
\textbf{Notes:} All masses are in units of \Msun.
\end{minipage}
\end{center}
\end{table*}

Estimating stellar masses for an observed sample of galaxies involves obtaining the luminosities of the galaxies and then converting these to stellar masses
through a stellar population synthesis model and dust corrections. The errors introduced in all these steps are typically of the order of the uncertainties in the
stellar populations synthesis analysis.
Assuming that these errors are Gaussian, the shape of the SMF implies that
at a given observed mass there are more galaxies with a lower true mass than galaxies with a higher true mass, i.e. more galaxies are scattered in than
scattered out. As a result the observed SMF is shallower than the true SMF, particularly at the massive end. It is important to include this effect
in our model.

We do this by adding a Gaussian scatter to the stellar mass of each model galaxy. An \lq observed\rq~stellar mass in our model is thus obtained in
three steps: We first compute the average mass for a given halo mass and redshift (or infall halo mass and infall redshift for subhaloes), then we add a
fixed intrinsic scatter of $\sigma_m(\log m)=0.15$ dex to account for different halo parameters and histories, and finally we
add an observational scatter to account for errors in the observational estimate of stellar mass. The value of this scatter is $\sigma_{\rm obs}=0.1$ dex for the
SDSS SMF (as estimated by LW09), $\sigma_{\rm obs}=0.3$ dex for the SMFs by PG08 and S12 at $z<3$ and
$\sigma_{\rm obs}=0.45$ dex for the SMFs at $z\ge3$ (see appendix B of PG08).

\subsubsection{The evolution of the SHM parameters}
\label{sec:ome_matching}

First we employ the classical abundance matching approach as described in Section \ref{sec:matching}, now taking into account observational mass errors, to
investigate how the SHM parameters evolve with redshift, and if a different functional form has to be adopted. The results are plotted in Figure
\ref{fig:parametersS} (symbols with errorbars). As for the fit without observational mass errors, the transition mass $M_1$ increases with increasing redshift.
This increase is strong at low redshift, while at high redshift $M_1$ increases only slightly. Unlike before we do not find that $M_1$ has
a maximum beyond which it decreases again. This result implies that at high redshift, massive haloes with $\log(\Mvir/\Msun)\sim12.5$ 
are most efficient in converting baryons into stars, while at lower redshift the halo mass for which the baryon conversion efficiency is
maximal, decreases towards $\log(\Mvir/\Msun)\sim11.6$.

The normalization $N$ (the conversion efficiency at the peak) is now found
to decrease monotonically with redshift. The reason why the normalization does not change much at
low redshift with respect to the model without observational scatter, but is very different at high redshift, is the evolution of the low-mass
slope $\alpha$ of the SMF. At low redshift this slope is very shallow, so relatively few galaxies are scattered from lower masses towards
the characteristic mass of the SMF (which corresponds to the peak efficiency). At high redshift, the low-mass
slope is much steeper, so that more galaxies are scattered towards (and beyond) the mass of the peak efficiency. As a result, the
observed SMFs at low redshift are very similar to the true SMFs, while at high redshift the observed SMFs are shifted towards higher
masses. Therefore it appears as if the normalization is increasing at high redshift, if the observational mass error is not taken into
account. When we account for this, we find that the true SMFs are shifted to lower masses and the normalization of the SHM ratio $N$
no longer increases at high redshift.

The results for the low-mass slope $\beta$ are very similar to what we found before. This is a consequence of the shallow low-mass
slope of the SMFs, so that the observational mass error does not change the SMFs much. As before, $\beta$ is decreasing with
increasing redshift, so that the slope of the SHM relation becomes shallower with increasing redshift. Again, $\beta$ is only constrained by the
S12 SMFs.

\begin{figure*}
\psfig{figure=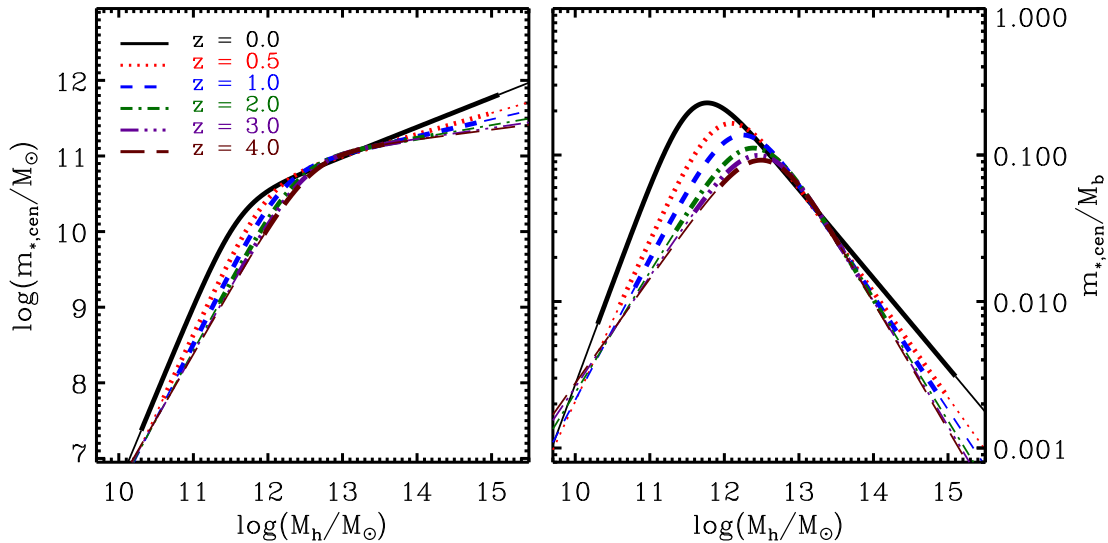,width=0.9\textwidth}
\caption{Stellar-to-Halo mass relation for central galaxies as a function of redshift. This relation is also valid for satellite galaxies when
virial mass and current redshift are substituted with subhalo infall mass and infall redshift, so that the satellite stellar mass is obtained
when it was last a central galaxy and does not change until it is accreted onto the central galaxy. \textit{Left panel:} Evolution of the SHM
relation. \textit{Right panel:} Evolution of the baryon conversion efficiency $m_{\rm *,cen}/M_b=m_{\rm *,cen}/(\Mvir f_b)=(m_{\rm *,cen}/\Mvir)
\times(\Omega_m/\Omega_b)$. The thick lines correspond to the ranges where the relation is constrained by data, while the
thin lines are extrapolations.}
\label{fig:sheffS}
\end{figure*}

The high-mass slope $\gamma$ is very different to what was found without observational mass errors. Where before $\gamma$ was
decreasing with increasing redshift, we now find that $\gamma$ is increasing. The reason for this is the steep slope of the SMFs at the massive
end, which means that the true SMFs are even steeper than the observed SMFs. Therefore the high mass slope of the SHM relation has to
become shallower which yields a larger $\gamma$. Here, the constraints are provided only by the PG08 SMFs, as only this survey
is wide enough to detect very massive galaxies.

\subsubsection{The multi-epoch abundance matching model including observational mass errors}
\label{sec:ome_asm}

Now that we have checked how the SHM parameters evolve with redshift, we can again employ our self-consistent model. As this evolution is
found to be different with respect to the model without observational mass errors, we first need to determine the new redshift parameterization
of the SHM parameters. As all parameters evolve considerably at low redshift, but evolve only slightly at high redshift, we found it more convenient to
use a linear dependence on the expansion parameter $a=(z+1)^{-1}$:
\begin{eqnarray}
\log M_1(z) &=& M_{10}+M_{11}(1-a)=M_{10}+M_{11}\frac{z}{z+1},\\
N(z) &=& N_{10}+N_{11}(1-a)=N_{10}+N_{11}\frac{z}{z+1},\\
\beta(z) &=& \beta_{10}+\beta_{11}(1-a)=\beta_{10}+\beta_{11}\frac{z}{z+1},\\
\gamma(z) &=& \gamma_{10}+\gamma_{11}(1-a)=\gamma_{10}+\gamma_{11}\frac{z}{z+1}.
\end{eqnarray}

The stellar mass is given by equation \eqref{eqn:shmratio} as a function of the virial mass of the main halo \Mvir and redshift $z$ for central galaxies
and as a function of the infall mass of the subhalo $M_{\rm inf}$ and infall redshift $z_{\rm inf}$ for satellite galaxies (see equation
\eqref{eqn:censat}). We constrain the eight parameters $M_{10}$, $M_{11}$, $N_{10}$, $N_{11}$, $\beta_{10}$, $\beta_{11}$, $\gamma_{10}$
and $\gamma_{11}$ by fitting our model to a set of SMFs at different redshifts. Again we employ the simulation at those redshifts where
an observed SMF is available and populate all main haloes and subhaloes in the simulation with galaxies using the redshift-dependent
SHM relation to derive their stellar masses. We then compute the 19 model SMFs and employ an MCMC method to find the best-fit values and
their $1\sigma$ errors. The results are summarized in Table \ref{t:pars}.

We show the results in Figure \ref{fig:parametersS} for the simultaneous fit to all SMFs (black line and shaded area). The redshift evolution of all four SHM
parameters in the simultaneous fit follow the trends that have been seen for the fits at individual redshifts. 
We plot the SHM relation and the baryon conversion efficiency for a set of redshifts between $z=0$ and $z=4$ in Figure \ref{fig:sheffS}.
At a fixed halo mass, the stellar mass now decreases with increasing redshift for low mass, intermediate and massive
galaxies. As a consequence, all central galaxies grow with time.
Moreover, independent of halo mass, we find that in all systems 
the central galaxies grow faster in stellar mass than the respective main haloes accrete dark matter.

\begin{figure*}
\psfig{figure=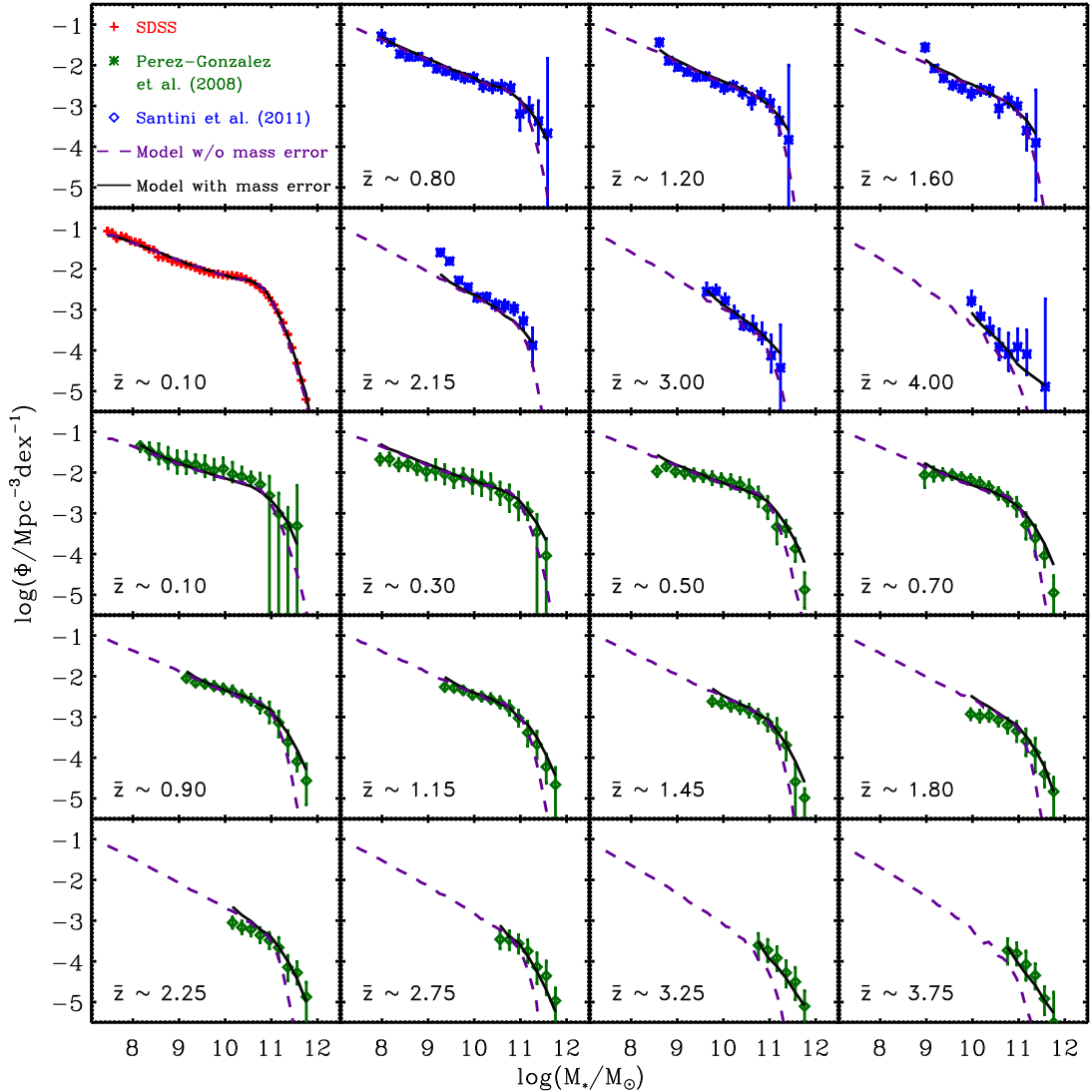,width=0.85\textwidth}
\caption{Comparison of the model (lines) and the observed SMFs (symbols). The redshift is indicated in each panel. The model SMFs have been computed
using the results of the fit including observational mass error. The solid lines show the SMFs when mass errors are included and represent a fit to the observations.
The dashed lines show the \lq true\rq~SMFs, before observational mass error is added.}
\label{fig:smfS}
\end{figure*}

Finally we compare our model SMFs to the observed ones in Figure \ref{fig:smfS}. The model SMFs have been computed using the result of the simultaneous fit
to all SMFs, i.e. the parameters given in Table \ref{t:pars}. The dashed lines show the \lq true\rq~SMFs, that have been computed after intrinsic scatter has been
added to the galaxy catalogue, but before observational mass error. The solid lines show the model SMFs, after mass errors have been included. At low redshift
and at low stellar masses, the observational mass errors do not impact the shape of the SMFs significantly, as the slope of the SMF is
shallower than at higher redshift and at the massive end. Beyond the characteristic stellar mass, the \lq true\rq~SMFs are much steeper than the observed ones,
so that there are substantially fewer very massive galaxies at high redshift than the observational SMFs appear to indicate.

The model SMFs including observational mass error are in very good agreement with the observed SMFs. This indicates, that our parameterization of the redshift
evolution of the SHM relation is compatible with observational constraints. However, there are a number of discrepancies, that stem mainly from the fact that the
observed SMFs of different authors are not in agreement. The SMF of PG08 at low redshift ($z=0.1$) is higher than our model, while the SDSS
SMF is fit very well. The reason for this is that the SDSS has much smaller errors than the PG08 SMF, so that during the fit a much higher likelihood is
given to the model that agrees with the SDSS. At higher redshift there are discrepancies between the model and the observed SMFs at the low-mass end. While
the model overpredicts the abundance of low-mass galaxies with respect to the PG08 SMFs, it underpredicts the abundance with respect to the S12 SMFs. This is
due to the PG08 SMFs having a much shallower slope than the S12 SMFs. As the errors on both sets of observations are of the same order, a model is favored that
yields average abundances.

\section{The evolution of the stellar content}
\label{sec:growth}

\begin{figure*}
\psfig{figure=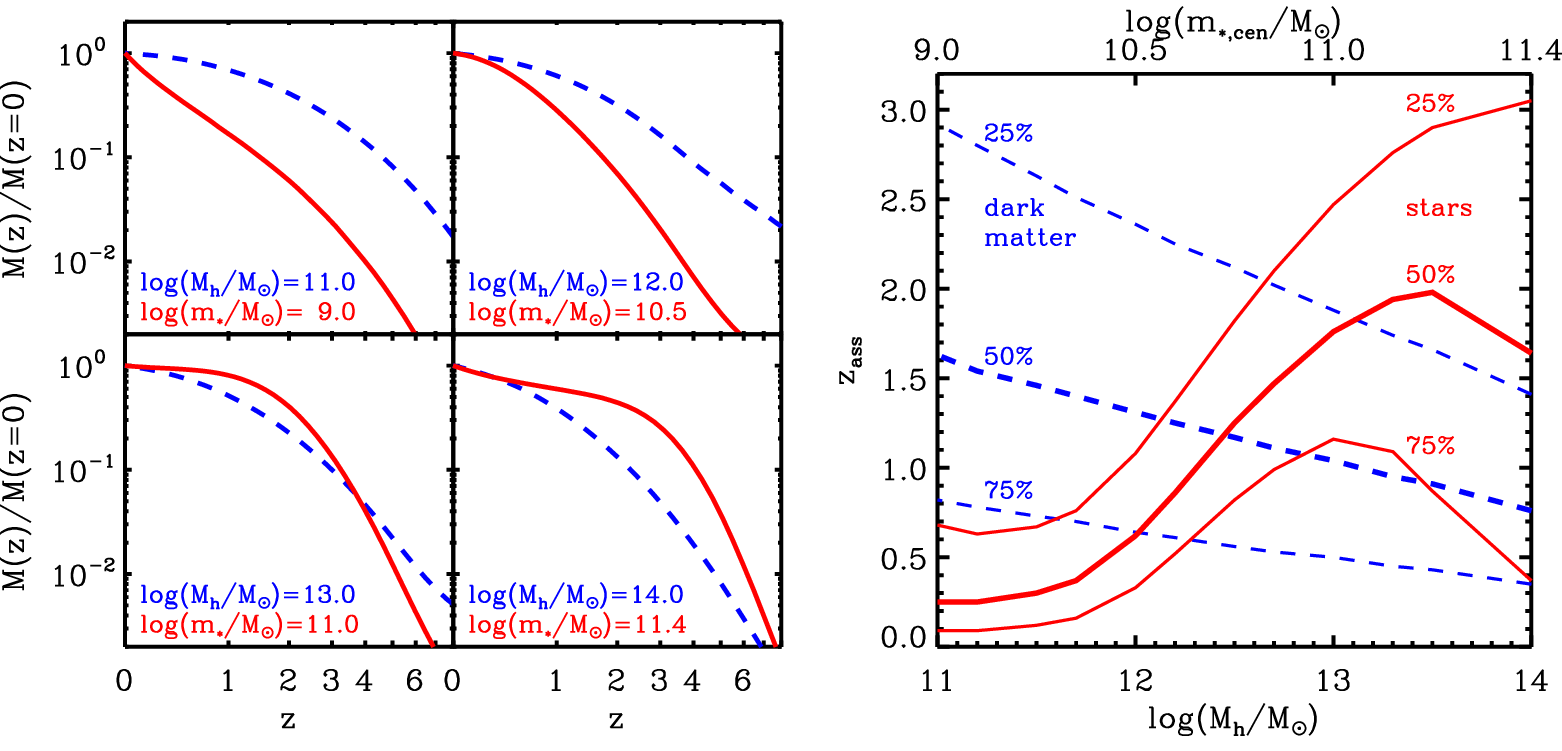,width=0.95\textwidth}
\caption{\textit{Left panels:} Average fraction of $z=0$ mass assembled as a function of redshift for dark matter haloes (blue dashed lines) and central galaxies
(red solid lines). Each panel compares the mass assembly history of a central galaxy to that of its parent halo. The $z=0$ halo and stellar masses are
indicated in each panel. While for low-mass dark matter haloes most of the mass assembles at early times, massive haloes only assemble late.
For galaxies these trends are opposite. \textit{Right panel:} Average formation redshift of dark matter haloes (blue dashed lines) and central galaxies
(red solid lines) as a function of $z=0$ mass. The three different lines indicate the redshift at which 25, 50, and 75 per cent of the mass was in place.}
\label{fig:assembly}
\end{figure*}

Having constrained the evolution of the SHM relation for central galaxies, we can now use our model to infer the evolution of the stellar content of galaxies.
To this end we employ the dark matter halo merger trees that have been extracted from the \textit{MS} and \textit{MS-II}. These merger trees connect haloes at
any given epoch to their descendants at later epochs or to their progenitors at earlier epochs. As our redshift-dependent SHM relation determines the
connection between galaxies and haloes at any epoch, we can use the connection between haloes across time to infer the connection between galaxies
across time. By averaging over a sample of haloes with the same mass at $z=0$, our model yields the average stellar mass growth of central galaxies as a
function of halo mass. We apply this method for sets of halo masses between $10^{11}\Msun$ and $10^{14}\Msun$.
This analysis is related to previous studies that employed semi-analytic models \citep{delucia2006,delucia2007,guo2008}. For a comparison and
discussion we refer to Section \ref{sec:disc}.

As a first step, we select a sample of merger trees by fixing the virial mass of the main halo at $z=0$. In the next step, we populate the haloes in these merger
trees using our redshift-dependent SHM relation to compute the stellar masses of their central galaxies. At every epoch, i.e. simulation snapshot, we then
compute the average stellar mass of the central galaxies in the main haloes. Finally, we calculate the difference of the average central galaxy stellar mass
between two subsequent snapshots and divide this by the time that has elapsed between those snapshots to derive the average stellar mass growth rate of
the central galaxies.  By doing this for a set of trees in halo mass bins, we can compute the average growth rate as a function of the $z=0$ halo mass and redshift.

In the left panels of Figure \ref{fig:assembly} we show the average fraction of $z=0$ mass assembled as a function of redshift for dark matter haloes and
central galaxies. For four $z=0$ halo masses, the assembly history of a central galaxy is compared to that of its parent halo. In low-mass systems with
$\log(\Mvir/\Msun)\lta12.5$ the halo grows much slower than its central galaxy. This indicates that in these systems the growth is dominated by internal
gas physical processes such as cooling and star formation. Only rather little external material is falling into the halo. For massive systems with
$\log(\Mvir/\Msun)\gta12.5$ the trends are opposite at low redshift: the fractional growth of haloes since $z\sim2$ is much stronger than the stellar growth
of their central galaxy. A large fraction of the stellar mass of massive galaxies is already in place at $z\lta2$. In these massive systems the growth is
dominated by external processes at late times, as they are accreting large amounts of material. While the halo grows rapidly, the central galaxy does not,
as most of the accreted systems take a very long time to reach the centre of the halo and instead survive to $z=0$ as \lq satellite\rq~ members of the group
or cluster.

To further study when most of the stellar and halo mass was in place we follow and extend the analysis by \citet{conroy2009}.
We plot the average assembly redshift of main haloes and central galaxies 
as a function of $z=0$ mass in the right panel of Figure \ref{fig:assembly}. We show the redshift at which 25, 50, and 75 per cent of the present stellar and
halo mass was already in place. In low-mass systems, the stellar mass assembled much later than the halo mass. For a system with
$\Mvir=10^{12}\Msun$, half of the stellar mass of the central galaxy forms after redshift $z=0.7$ while half of the virial mass was already assembled
by redshift $z=1.3$. In more massive systems haloes assemble later while galaxies assemble earlier: for a system with a $z=0$ halo mass of
$\Mvir=10^{13}\Msun$, half of the halo has assembled only by redshift $z=1.0$, but half of the stellar mass was already in place by redshift $z=1.8$.

Very massive galaxies in haloes with $\Mvir=10^{14}\Msun$ show a very interesting behavior: while they begin to form very early and have 25 per cent of their
stellar mass in place already by redshift $z=3.0$, it takes them a long time to fully assemble. Half of their stellar mass is in place only by redshift $z=1.6$,
while 75 per cent is assembled by $z=0.3$. The final $z=0$ stellar mass is then assembled very quickly. The reason why these massive galaxies grow very
fast at early times, only grow slowly at intermediate redshifts and grow fast again at late times can be found in the two processes that contribute
to the growth of galaxies. At high redshift, a massive galaxy grows effectively by star formation, while at low redshift, stellar accretion leads to a fast growth.
At intermediate redshift however, neither of these processes is effective, so that the galaxy grows only very slowly. In the following section we aim to constrain
and separate these two processes. Note that while massive galaxies assemble their final stellar mass only late, the stars in these galaxies have formed
much earlier, such that this result is not in conflict with the observation that the stars in massive systems are older than stars in low-mass galaxies.

\subsection{Star formation and accretion histories}
\label{sec:sfr}

The stellar mass of a galaxy can grow either by star formation inside the galaxy (in-situ) or by the assembly of stars that formed outside the galaxy and have
been accreted later (ex-situ). In principle, both processes can contribute simultaneously to the growth of a central galaxy, but in practice one process will dominate
over the other as a function of halo mass and redshift. In order to separate the contributions from star formation and accretion, we have first to determine the
amount of accretion of subhaloes onto the central object in the halo. Using our model, we can then populate these subhaloes with satellite galaxies and
compute the stellar accretion rate. Finally, the SFR can be determined from the difference between the total growth rate and the accretion rate.

We use the same set of dark matter halo merger trees as before, i.e. we use all trees with a common $z=0$ virial mass of the main halo and assign a stellar
mass to all haloes and subhaloes using our redshift-dependent SHM relation. For every simulation snapshot, we then record all satellite galaxies that are
accreted onto the central galaxy until the time of the next snapshot. We note, that, by construction, this also includes orphan galaxies. At each snapshot we
then sum the stellar mass of all recorded merging galaxies and divide this by the number of central galaxies in the sample (which corresponds to the number
of merger trees). We further assume that a fraction of satellite stars $f_{\rm esc}=0.2$ escapes from the central galaxy and ends up in the halo as diffuse stellar
material not detected in standard surveys.
This fiducial value is motivated by high-resolution simulations of binary galaxy mergers \citep[e.g.][]{nipoti2003,naab2003,moster2012}. For a different value the
resulting stellar accretion and star formation rates change accordingly (e.g. for $f_{\rm esc}=0$ the accretion rates are 25 per cent higher).
In this way we get the average stellar accretion rate onto the central galaxy as a function of the $z=0$ halo mass and redshift.

\begin{figure*}
\psfig{figure=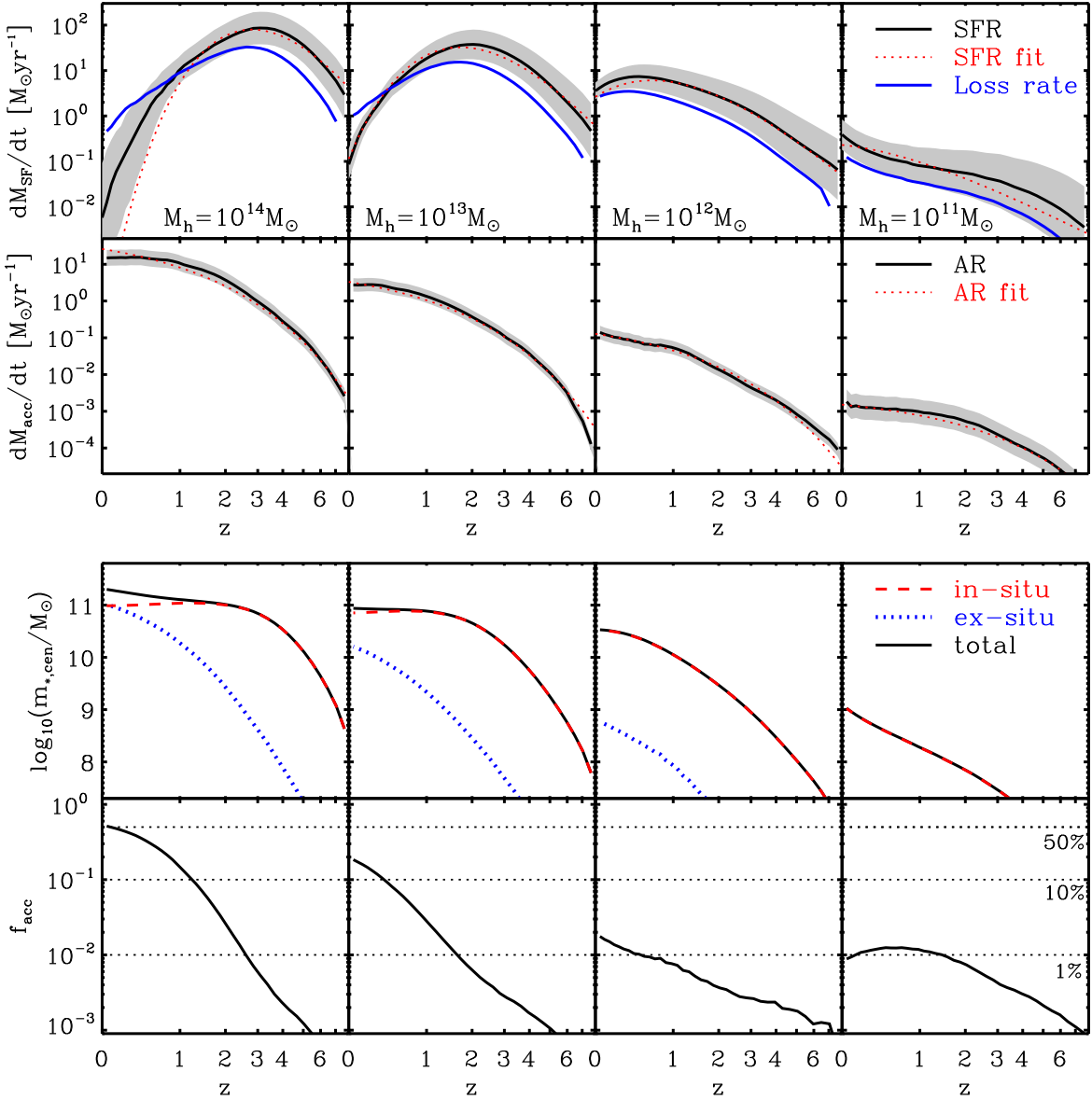,width=0.9\textwidth}
\caption{Star formation and accretion histories for different $z=0$ halo masses. The rows from top to bottom show the star formation history, the stellar accretion
history, the stellar mass formed in-situ and ex-situ, and the fraction of ex-situ formed stars. Each column represents a different $z=0$ halo mass. In the first two
rows the black lines and shaded areas represent the rates that have been derived with our model and their $1\sigma$ confidence levels. The red lines are fits
to these relation as given by equations \eqref{eqn:sfrfit}-\eqref{eqn:sfrpars} and \eqref{eqn:accfit}-\eqref{eqn:accpars}. The blue line in the first row shows the
stellar mass loss rate. In the third row, the red dotted lines give the stellar mass of in-situ formed stars that have survived, the blue dashed lines give the
stellar mass that has formed ex-situ, and the black line is the total stellar mass in the central galaxy. The dotted lines in the fourth row indicate that 50, 10 and
1 per cent of the stellar mass has formed ex-situ.}
\label{fig:accretion}
\end{figure*}

Having determined the total stellar growth rate and the amount of growth that is due to accretion, the SFR of an average galaxy can be computed from the
residual stellar mass growth attributed to star formation. However, the relation between this residual growth rate and the SFR is complicated by mass loss
due to dying stars:
\begin{equation}
\label{eqn:rates}
\frac{{\rm d}m_{\rm res}}{{\rm d}t} = \frac{{\rm d}m_{\rm tot}}{{\rm d}t} - \frac{{\rm d}m_{\rm acc}}{{\rm d}t} = \frac{{\rm d}m_{\rm SF}}{{\rm d}t}
- \frac{{\rm d}m_{\rm loss}}{{\rm d}t} = \frac{{\rm d}m_{\rm IS}}{{\rm d}t}\;.
\end{equation}
This equation states that the residual growth rate ${\rm d}m_{\rm res}/{\rm d}t$, which is defined as the difference between the total growth rate
${\rm d}m_{\rm tot}/{\rm d}t$ and the accretion rate ${\rm d}m_{\rm acc}/{\rm d}t$, is given by the difference between the SFR ${\rm d}m_{\rm SF}/{\rm d}t$
and the stellar mass loss rate ${\rm d}m_{\rm loss}/{\rm d}t$, and corresponds to a net in-situ growth rate ${\rm d}m_{\rm IS}/{\rm d}t$. It can be solved
iteratively, starting with a SFR that is equal to the residual growth rate at every epoch. We go to the time of the first snapshot and compute the amount of stellar
mass that is formed in an average galaxy. Then we move to the second snapshot and compute both the stellar mass that forms at this time and the stellar mass
that is lost in the population that formed in the first snapshot. For the latter we employ the formula
\begin{equation}
f_{\rm loss} (t) = 0.05 \times \ln\left(\frac{t+3\times10^5 {\rm~yr}}{3\times10^5 {\rm~yr}}\right) \;,
\end{equation}
where $f_{\rm loss} (t)$ is the fraction of mass lost by a stellar population with an age $t$. This formula is a fit to the mass loss of a stellar population with
a \citet{chabrier2003} IMF as developed by \citet{bruzual2003} and implies that 40 per cent of stellar mass is lost within $1\Gyr$ and 50 per cent of stellar mass
is lost within $10\Gyr$.

We then move to the third snapshot, compute the stellar mass that forms at this time and derive the stellar mass that is lost in the populations
that formed in the first and second snapshots. This procedure is continued until $z=0$, while at every snapshot the amount of lost stellar mass is the sum of the
lost mass of all populations that have formed until then. We can then compute the loss rate between subsequent snapshots and the in-situ growth rate as the
difference between SFR and loss rate. This in-situ growth rate is then compared to the residual growth rate as described by equation \ref{eqn:rates}. 
The difference between those rates is then added to the SFR (separately at each snapshot) and the procedure is repeated and iterated until the in-situ growth
rate and the residual growth rate are equal. We repeat this procedure for each sample of halo merger trees with a given $z=0$ halo mass.

The resulting SFRs and accretion rates are shown in the first and second rows of Figure \ref{fig:accretion} (black solid lines), respectively, for four different
$z=0$ halo masses. The $1\sigma$ confidence levels for both rates (shaded areas) have been computed by employing the upper and lower estimates for
mean stellar masses according to our SHM relation. We have also added the stellar mass loss rates in upper panels (blue line). All results represent the average
rates, i.e. the average of both red and blue galaxies at a given $z=0$ halo mass. Of course, individual galaxies can have higher or lower SFRs depending
on other galaxy properties like color, so that blue (red) galaxies can have higher (lower) SFRs.

We find that the star formation history of central galaxies peaks at a certain redshift. This peak SFR redshift depends on halo mass: for massive systems this
peak occurs at high redshift ($z=3$ for $\Mvir=10^{14}\Msun$) while for lower masses the SFR peaks at a lower redshift ($z=0.5$ for $\Mvir=10^{12}\Msun$).
For systems with an even lower mass of $\Mvir\lta10^{11.7}\Msun$ the peak has not been reached yet by $z=0$. After the peak has been reached, the SFR
quickly drops, such that massive galaxies are quenched at $z\sim2$. Furthermore, the peak SFR is higher in more massive systems. We also find that in systems
that are more massive than $\Mvir=10^{12.5}\Msun$, the stellar mass loss rate due to dying stars is larger than the SFR at $z=0$, such that the mass of stars
present in the central galaxy is decreasing.

The accretion rates increase with time for all systems. Massive systems have a high accretion rate that dominates the total growth rate at late times, while for
low-mass systems, accretion is insignificant at all times. For a system with $\Mvir=10^{14}\Msun$ the accretion rate dominates over the SFR at $z<1$. The 
halo mass at which the accretion rate and the SFR are equal at $z=0$ is $\Mvir=10^{12.7}\Msun$. For lower masses the contribution of accretion to the total
growth rate quickly drops. The reason why accretion contributes significantly to the overall growth in massive systems and it is insignificant in low-mass systems
can be inferred from the SHM relation. Massive haloes host relatively massive subhaloes, and these subhaloes are filled with relatively massive satellite galaxies.
The subhaloes that have e.g. one tenth of the main halo mass host galaxies that are almost as massive as the central galaxy, so that the resulting accretion rates
are high. In low-mass systems, however, the subhaloes are almost entirely devoid of stars, so that the accretion of the subhaloes onto the center does not result
in a significant stellar accretion rate. Moreover, Figure \ref{fig:assembly} shows that low-mass haloes have largely assembled by $z=1$, so that any late time
growth has to come from star formation.

To compare further the relative importance of star formation and accretion for systems of different mass, we plot in the third row of Figure \ref{fig:accretion}
the amount of stellar mass that has formed in-situ, i.e. through star formation, and the amount that has formed ex-situ, i.e. has been accreted. In the fourth row we
show the fraction of stellar mass of the central galaxy that has formed ex-situ. In massive systems with $\Mvir \ge 10^{13}\Msun$, all of the stars that have formed
in-situ and survived were already in place at high redshift ($z\gta2$). Although, the loss rate is higher than the accretion rate at late times, and the amount of in-situ
stellar mass has decreased since $z\sim2$, the total stellar mass of the system increases with time, as the accretion rate compensates for the lost stellar mass.
In very massive systems with $\Mvir=10^{14}\Msun$ more than half of the central stellar mass is formed ex-situ. In lower mass systems the amount of ex-situ
stellar mass is decreasing considerably: for $\Mvir=10^{13}\Msun$ the fraction of ex-situ stars is 20 per cent, for $\Mvir=10^{12}\Msun$ the fraction is
2 per cent, and for $\Mvir=10^{12}\Msun$ the fraction is 1 per cent. At high redshift ($z\gta2$), the amount of stellar mass that has formed ex-situ is negligible for
systems of all masses. We further note that the accretion rates grow more rapidly in massive systems than in systems of low mass.

The connection between star formation and accretion history and the virial mass of the system can be of general use in constraining models of cooling, 
star formation and feedback in galaxies. These relations have been derived entirely from observations with an empirical model. Although this model makes
no reference to the physical processes that drive the evolution of galaxies, it can be very helpful in order to constrain more physical models. In order to easily
compare our model predictions to the results of other models such as simulations and semi-analytic models, we provide simple fitting functions to the
star formation and accretion histories.

We find that at a given $z=0$ main halo mass the SFR is well fit by the relation:
\begin{eqnarray}
\label{eqn:sfrfit}
\frac{{\rm d}m_{\rm SF}}{{\rm d}t} (z) &=& f_1 \; a^{f_2} \; \exp\left(\frac{f_2}{f_3}\;\left(1.0-a\right)\right) \nonumber\\
&=& f_1 \left(1+z\right)^{-f_2} \exp\left(\frac{f_2}{f_3}\frac{z}{1+z}\right).
\end{eqnarray}
The parameters $f_1$, $f_2$, and $f_3$ depend on the $z=0$ halo mass of the system and follow the relations:
\begin{eqnarray}
f_1 &=& f_{10} \; \exp \left(-\frac{(\log(\Mvir/\Msun)-f_{11})^2}{2 \; f_{12}^2}\right)\;,\\
f_2 &=& f_{20}+f_{21} \; \log\left(\frac{\Mvir}{10^{12}\Msun}\right) \;,\\
\log(f_3) &=& f_{30}+f_{31} \left(\frac{\Mvir}{10^{12}\Msun}\right)^{f_{32}}\;.
\label{eqn:sfrpars}
\end{eqnarray}
The parameter $f_1$ can be interpreted as the SFR at $z=0$, $f_2$ is the slope of the star formation history at high redshift, and $f_3$ corresponds to the
expansion factor $a$ of the peak SFR. The values of the fitting parameters are:\\
\null\\
\begin{tabular}{@{}lll@{}}
$f_{10} = 2.658~\Msunpyr$, & $f_{11} = 11.937$, & $f_{12} = 0.424$,\\
$f_{20} = 5.507$, & $f_{21} = 2.437$, & \\
$f_{30} = -0.915$, & $f_{31} = 0.696$, & $f_{32} = -0.159\;.$\\
\end{tabular}
\null\\

\noindent
The SFRs obtained with this fitting function are compared to the inferred ones in Figure \ref{fig:accretion} (red lines). Overall the global fit is very close to the model
predictions. For very massive galaxies with $\Mvir=10^{14}\Msun$ at low redshift ($z<1$), the fitting function is lower than the model SFR, although it is still within
the $1\sigma$ confidence levels.

The accretion history for systems of a given $z=0$ main halo mass can be described with an exponential dependence on redshift:
\begin{eqnarray}
\label{eqn:accfit}
\frac{{\rm d}m_{\rm acc}}{{\rm d}t} (z) = g_1 \; \exp\left(-\frac{z}{g_2}\right)\;.
\end{eqnarray}
The parameters $g_1$, and $g_2$ are functions of the $z=0$ halo mass of the system and are given by:
\begin{eqnarray}
g_1&=& g_{10} \left[\left(\frac{\Mvir}{10^{12.5}\Msun}\right)^{g_{11}}+\left(\frac{\Mvir}{10^{12.5}\Msun}\right)^{g_{12}}\right]^{-1},\\
g_2 &=& g_{20}+g_{21} \left(\frac{\Mvir}{10^{12}\Msun}\right)^{g_{22}}.
\label{eqn:accpars}
\end{eqnarray}
The parameter $g_1$ gives the stellar accretion rate at $z=0$, and $g_2$ is the slope of the accretion history. The values of the fitting parameters are:\\
\null\\
\begin{tabular}{@{}lll@{}}
$g_{10} = 1.633~\Msunpyr$, & $g_{11} = -2.017$, & $g_{12} = -0.806$,\\
$g_{20} = 0.855$, & $g_{21} = 0.098$, & $g_{22} = -0.797\;.$\\
\end{tabular}
\null\\

\noindent
The accretion rates obtained with this fitting function are compared to the inferred ones in Figure \ref{fig:accretion} (red lines). The global fit is very close to the model
prŽdictions for all redshifts and halo masses.

\subsection{Cosmic star formation rate density}

\begin{figure}
\psfig{figure=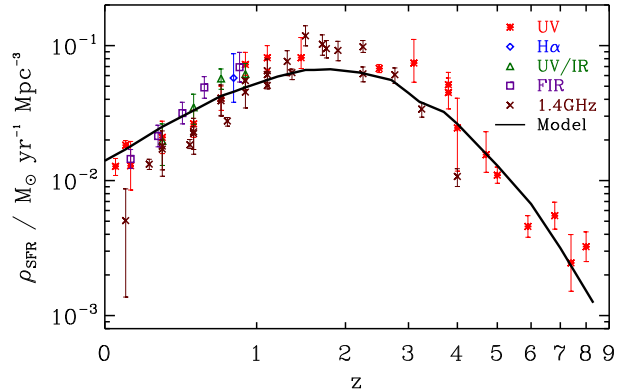,width=0.45\textwidth}
\caption{Star formation rate density as a function of redshift. The symbols of different colors give observational estimates derived with different methods. The black line
is the result of our model. The model was constrained with observed SMFs at different redshifts and has not been fit to observed SFRs. Still the model is in excellent
agreement with the data.}
\label{fig:sfrd}
\end{figure}

In order to check whether our model SFRs are in agreement with observational constraints, we compare our model prediction for the evolution of the cosmic SFR
density to data in the literature. For this we compute the SFR for each halo in the simulations using our recipe as given by equations
\eqref{eqn:sfrfit}-\eqref{eqn:sfrpars} and compute the volume density for each snapshot. We compare the result to observational estimates derived with different
methods in Figure \ref{fig:sfrd}. This comparison includes estimates derived from the UV \citep{salim2007,vdburg2010,bouwens2009,robotham2011,bouwens2011,
cucciati2012}, ${\rm H}\alpha$ \citep{ly2011}, both UV and IR \citep{zheng2007}, the FIR \citep{rujopakarn2010} and from radio observations \citep{smolcic2009,
dunne2009,karim2011}.

Our model is in excellent agreement with the data at all redshifts. We note that our model was only constrained with observed SMFs at different redshifts and has not
been fit to observed SFRs. This means that the observed SMFs are broadly consistent with observational inferences for SFRs once observational mass errors
are taken into account. Moreover, this shows that our choice for the parameterization for the redshift-dependent SHM relation is consistent with observational
constraints, and that our inferred recipe for the mass-dependent star formation histories gives reliable estimates.

\subsection{Specific star formation rates}

\begin{figure}
\psfig{figure=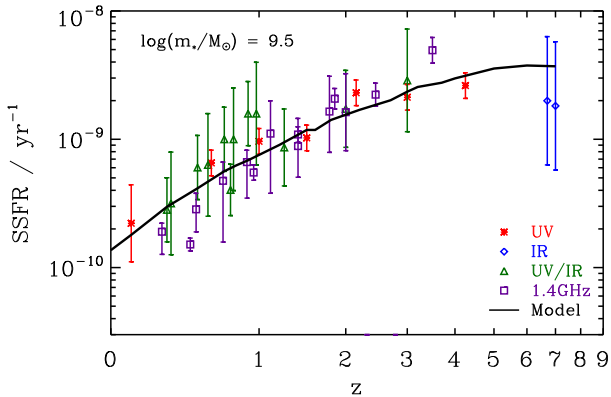,width=0.45\textwidth}
\caption{Specific star formation rate for galaxies with a stellar mass of $\log(m_*/\Mvir)=9.5$ at the various epochs. The symbols of different colors give observational
estimates derived with different methods and the black line is the result of our model. Although this relation has not been fit, the model is in excellent agreement with
the data.}
\label{fig:ssfrA}
\end{figure}

\begin{figure}
\psfig{figure=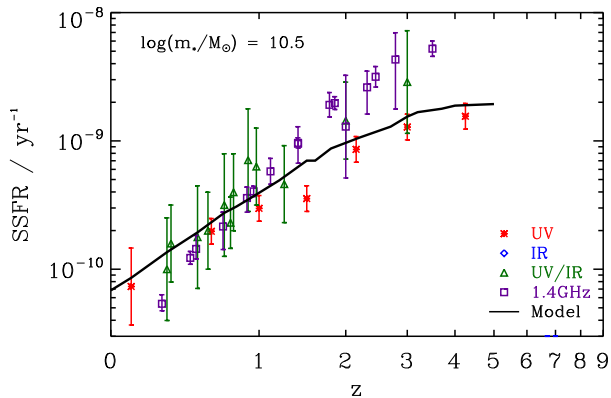,width=0.45\textwidth}
\caption{Same as Figure \ref{fig:ssfrA}, but for galaxies with a stellar mass of $\log(m_*/\Mvir)=10.5$ at the various epochs.}
\label{fig:ssfrB}
\end{figure}

As a further test whether our model predictions are in agreement within the observational limits, we compute the specific star formation rate (SSFR) for galaxies of
a fixed stellar mass. Observationally, it is very difficult to infer the SSFR as a function of the $z=0$ stellar mass since observations cannot connect galaxies at
different epochs. Therefore it is not possible to follow the star formation history of an average galaxy across time with observations up to high redshift. However,
it is possible to derive the average SSFR for galaxies within a stellar mass range at different redshifts. We can compare these observational data to the predictions
of our model, by populating all haloes in a given snapshot with galaxies using our SHM relation to get stellar masses and our recipe for the star formation history
to get their SFR. We show this comparison for low-mass galaxies with $\log(m_*/\Mvir)=9.5$ in Figure \ref{fig:ssfrA} and for massive galaxies $\log(m_*/\Mvir)=10.5$
in Figure \ref{fig:ssfrB}. The comparison includes estimates derived from the UV \citep{feulner2005,salim2007}, the IR \citep{labbe2010,schaerer2010,mclure2011},
both UV and IR \citep{zheng2007,noeske2007,kajisawa2010}, and from radio observations \citep{daddi2007,dunne2009,karim2011}.

For both stellar mass ranges, our model is in excellent agreement with the data, although this relation has not been fit. This shows that mass-dependent quantities,
such as the SSFR can be predicted up to high redshift in addition to global quantities such as the cosmic SFR density. As our model is able to fit a wide range of
observed data, we are confident that it can be used to estimate reliably relations that cannot be easily observed, such as the star formation history of an average
galaxy as a function its $z=0$ halo mass.

\section{Discussion}
\label{sec:disc}

In this study we draw conclusions on the assembly histories of galaxies over a wide range in mass from a conceptually simple approach
matching observed galaxy mass functions and halo accretion histories from modern high resolution dark matter simulations which represent the
underlying cosmological model. The early domination of in-situ star formation and the growing importance of late accretion of ex-situ
stars for galaxies residing in more massive haloes found here is in good agreement with predictions from semi-analytical models
\citep{kauffmann1996,khochfar2006}. Furthermore, our analysis confirms the results of \citet{delucia2006} who find using semi-analytic models
that more massive (elliptical) galaxies have star formation histories that peak at higher redshifts than lower mass systems. 

Although star formation ceases at low redshift for massive systems, the stellar mass of their central galaxies can grow considerably through
accreted satellite galaxies. The majority of these ex-situ stars have formed when their galaxies still were centrals, i.e. before they entered the main
halo. Therefore massive galaxies consist of old stellar populations. On the other hand, massive galaxies start to assemble very early due to
star formation, but their final mass is assembled only very late. Employing semi-analytic models, \citet{delucia2007} find that half of the mass of
central galaxies of very massive haloes with $\Mvir=10^{15}\Msun$ forms after $z\sim4$, but is locked up in the central galaxy only after $z=0.5$.
This is in good agreement with our results: extrapolating the 50 per cent assembly redshift in Figure \ref{fig:assembly} to $\Mvir=10^{15}\Msun$
roughly yields $z=0.5$.

Our model indicates that low-mass (massive) galaxies form late (early), although their dark matter haloes assemble early (late). Although the haloes of
low-mass galaxies assembles early it takes a considerable amount of time for the gas to cool and fall to the centre and form stars. In massive systems,
on the other hand, the central galaxies (and the galaxies that will become satellites in these systems) can start to grow very early and form the bulk of
their stars at high redshift. However, their haloes assemble only late, and the satellites are accreted onto the central galaxies at low redshift. Therefore,
the stellar populations of massive galaxies are old, although massive haloes assemble late. This demonstrates that \lq downsizing\rq~is not anti-hierarchical,
but a consequence of the bottom-up $\Lambda$CDM scenario \citep{neistein2006,fontanot2009}.

The contribution of star formation and accretion to the total stellar growth of central galaxies depends both on halo mass and redshift. While at high
redshift accretion is negligible in all systems, at low redshift star formation dominates only in low-mass systems. The halo mass at which both processes
contribute equally to the total growth of the central galaxy at $z=0$ is $\log(\Mvir/\Msun)\sim12.7$, which corresponds to a stellar mass of
$\log(m/\Msun)\sim10.9$. Here, both modes add $\sim1\Msunpyr$ to the stellar mass of the galaxy. This is in good agreement with the results
of \citet{guo2008} who employ a semi-analytic model to study star formation and stellar accretion rates as a function of stellar mass and redshift.
At $z=0$ they find that the stellar mass of central galaxies for which both processes contribute equally is roughly $\log(m/\Msun)\sim10.6$, with rates
of $\sim0.5-1\Msunpyr$ (see their Figure 3). For systems of lower (higher) mass, both methods indicate a lower (higher) contribution of ex-situ stars.
At $z=2$ the semi-analytic model predicts that the stellar mass of galaxies with an equal contribution of the two processes is roughly
$\log(m/\Msun)\sim11.0$, with rates of $\sim20\Msunpyr$, while in our model only the most massive galaxies with $\log(m/\Msun)\gta11.4$ have accretion
rates that are comparable to their SFRs.

\begin{figure*}
\psfig{figure=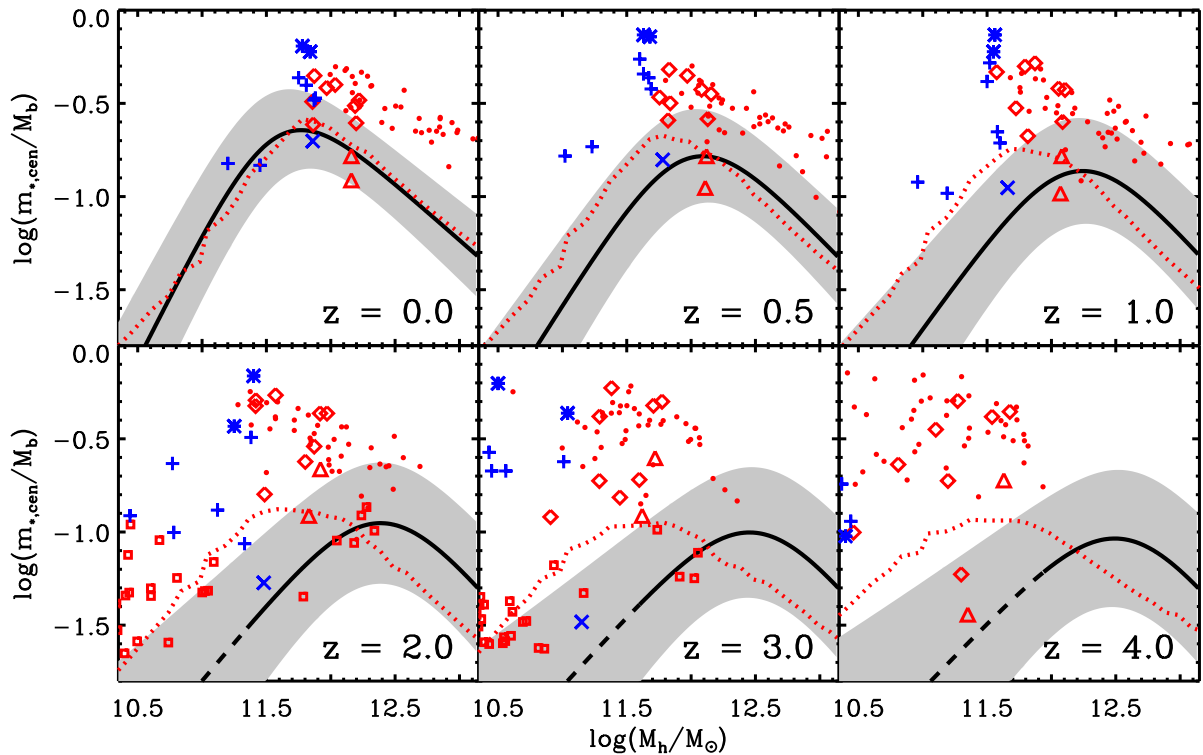,width=0.9\textwidth}
\caption{Comparison between central galaxy formation efficiencies found in numerical simulations at different redshifts. Each panel corresponds
to the indicated redshift. The solid black lines give the average conversion efficiencies needed to fit the observed SMFs and the shaded areas
indicate the 1$\sigma$ confidence levels. The symbols show the results of hydrodynamical zoom-in simulations run with the {\sc Gasoline} code
(blue asterisks: \citealt{brooks2011}, pluses: \citealt{governato2012}, crosses: Stinson et al. in prep.) and the {\sc Gadget} code
(red dots: \citealt{oser2010}, diamonds: \citealt{scannapieco2011}, squares: \citealt{genel2012}, triangles: \citealt{okamoto2012}).
The colored lines show the conversion efficiencies predicted by the semi-analytic model by \citet[][red dotted line]{guo2011}.
While many simulations agree well with the predicted conversion efficiency at $z=0$, most have too high values at earlier epochs, indicating that
they form their stars too early.
}
\label{fig:efficiencyS}
\end{figure*}

In the present study on average 50 per cent of the stars in galaxies in massive haloes ($10^{14}\Msun$) at $z=0$ are made in external galaxies and accreted
at later epochs. However, in lower mass haloes, like that of the Milky Way ($10^{12}\Msun$), this fraction is only a few per cent, making stellar accretion
relatively unimportant for the evolution of the central galaxies. This in accord with standard formation and enrichment scenarios for disk galaxies
\citep{matteucci1989,mo1998,prantzos1998,naab2006,kauffmann2012}.

Similar trends are reported from three-dimensional cosmological hydrodynamical zoom-in simulations, however, with larger
fractions of accreted stars for galaxies of all masses. For the most massive galaxies this can be up to 80 per cent \citep{oser2010,feldmann2010}. Still, it
is likely that this fraction is reduced to 50 percent (in the same mass bin) by the inclusion of strong stellar feedback which reduces the
stellar content of the (typically lower mass) accreted galaxies \citep{hirschmann2012}. In zoom-in simulations individual stellar particles can directly be
followed and it has been shown that ex-situ stars predominantly assemble at larger radii \citep[e.g.][]{laporte2012}. In massive spheroidal galaxies, where
the fraction of ex-situ stars is high, this material plays a significant role in the structural evolution of the galaxies \citep{naab2009,hopkins2009,
feldmann2010,oser2010,oser2012}. In lower-mass disk-like galaxies the accreted stars contribute to the build-up of the outer stellar halo and can also directly
contribute on the order of 10 percent to the stellar mass of the disk \citep[e.g.][]{abadi2006,font2011,tissera2012}. Interestingly, these simulations predict ex-situ
fractions of more than $\sim20$ per cent (up to 40-50 per cent) which is an order of magnitude higher then found in our study.

Our model constrains the mass budget of cosmological galaxy formation simulations not only at the present day but also at higher redshift. We compare our model
to the results of hydrodynamical zoom-in simulations and semi-analytic models at six redshifts in Figure \ref{fig:efficiencyS}. While some older simulations have too high
conversion efficiencies at $z=0$, most of the newer simulations show good agreement with the efficiencies needed to reproduce the local observed SMFs.
However, almost all simulations have far too high efficiencies at earlier epochs. This indicates that these simulations form too many stars early ($z\sim3$), while at
late times ($z\lta2$) star formation is quenched too strongly, so that most of the stellar mass at $z=0$ is already present at high redshift. The lack of late star formation
in simulations could possibly explain the difficulties in getting prominent discs. Most stars are formed at times where processes leading to the formation of spheroids
are common, e.g. major mergers  or spin flips \citep{sales2012,okamoto2012}. If the majority of stars form after these epochs, a more prominent disc forms.

The results of the semi-analytic models agree very well with our relation at $z=0$. This is not surprising as these models were tuned to fit the local SMF.
At higher redshift however, their predictions differ from our model, indicating as noted in the original papers that they do not yield good fits to the observed high
redshift SMFs. While the normalization and the slopes are in good agreement with our relation, the halo mass where the efficiency peaks ($M_1$), moves to
lower masses with increasing redshift.

We note that a direct comparison between stellar masses measured in simulations and observed stellar masses is difficult. Especially at high redshift, it is
complicated to estimate stellar masses observationally, as many different effects have to be taken into account. Moreover, one has to
make a number of assumptions when fitting the spectral energy distributions (SEDs). An alternative way to compare stellar masses in models and observations is
by making mock images of simulated galaxies using a stellar populations synthesis code and then obtaining a \lq photometric\rq~stellar mass with an SED fit.
However there are two complications with this approach: First, it depends on the assumptions that are made to obtain the mock images (dust modeling, stellar
populations synthesis modeling). Second, the reduction of the mock images would have to be done in the same way as for the observed data, which
is non-trivial. Therefore we have decided to conduct a simpler comparison within the limits of the quoted observational errors.

The characteristic shape of the SHM relation is a consequence of several different physical processes, that prevent the gas in the dark matter halo from cooling
(or reheat cold gas) and forming stars. The contribution of each process varies for systems of different mass. In low-mass systems, feedback from
supernova-driven winds can be very effective as winds can easily expel large amounts of gas from haloes with low escape velocities
\citep[e.g.][]{larson1974,dekel1986,oppenheimer2008,puchwein2012}. Furthermore the amount of cooled gas is reduced by the presence of a UV and an X-ray
background \citep{navarro1997,gnedin2000,benson2002,hambrick2011} and by cosmic rays \citep{jubelgas2008,wadepuhl2011}. These effects have different
efficiencies for different halo masses, such that the low-mass slope of the SHM relation can change at the halo mass, where the efficiencies of the two processes
cross. However, with the current data it is very hard to find this transition mass, as the error estimates are even larger than the possible variations in the slope.

In massive haloes the star formation efficiency may be lowered by feedback from active galactic nuclei \citep[AGN][]{springel2005,croton2006,bower2006,
somerville2008,hambrick2011,teyssier2011,martizzi2012,puchwein2012} which can heat the gas in the halo and prevent it from cooling. Another effect
that may be important for massive galaxies is gravitational heating \citep{khochfar2008,johansson2009}. In hydrodynamical zoom-in simulations this effect is
naturally included and results in a trend of lower galaxy formation efficiencies for more massive haloes, although still at levels which are significantly too high
\citep{oser2010,hirschmann2012}. In this framework, the characteristic halo mass for which galaxy formation is most efficient is the mass where the
combination of the various physical processes that reduce the efficiency is minimal. For massive galaxies the amount of in-situ stars is affected largely by
processes that are effective for more massive systems (AGN feedback and gravitational heating), while the amount of ex-situ stars is strongly affected by
processes that are effective for systems of lower mass (such as SN feedback and an ionizing background radiation).

Although the only observational constraints of our model are the observed SMFs from $z=0$ to 4, our predictions for the evolution of the cosmic SFR 
density are in very good agreement with direct observation. This indicates that the evolution of the SMFs is consistent with the evolution of the cosmic
SFR density, although the integrated observed SFR density was previously found to be higher than the observed stellar mass density \citep[e.g.][]{hopkins2006}.
There are two effects that help to reconcile this tension. First, we employ the recent measurements by S12, who find a considerably steeper low-mass slope for
the SMFs. This results in a larger amount of stellar mass in the low-mass end and thus increases the observed stellar mass density. Second, we include the
effects of observational mass error in our model. As a consequence we find an evolution in the high-mass end of the \lq true\rq~SMFs, especially at low redshift.
When both effects are taken into account, the tension between the evolution of the stellar mass density and the integrated SFR density is considerably lowered.

\section{Summary}
\label{sec:sum}

In this paper, we present a galaxy evolution model that assumes a tight relation
between the stellar mass of a galaxy and the mass of the dark matter halo in which it resides.
With this approach we populate haloes with galaxies over the entire cosmic period from $z\sim4$
to the present, using the $N$-body Millennium simulations and observed SMFs. Unlike the classical
abundance matching technique our new method is self-consistent, taking into account that subhaloes
are accreted at earlier times and with higher masses than when they are observed. We employ a
redshift-dependent parameterization of the stellar-to-halo mass relation, and populate haloes
and subhaloes with galaxies requiring that the observed SMFs at different redshifts be reproduced simultaneously.

We show that when observational mass errors are not taken into account, no consistent stellar mass evolution
of massive galaxies can be obtained, as the central galaxy of a typical halo of mass $\Mvir=10^{14}\Msun$ would lose
$\sim25$ per cent of its stellar mass from $z=1$ to 0. When we include observational mass errors, this
tension vanishes and massive galaxies also have a physically consistent stellar mass evolution. 
We find that the halo mass where galaxy formation is most efficient is $\log(\Mvir/\Msun)\sim11.8$ at $z=0$ and
increases with increasing redshift to $\log(\Mvir/\Msun)\sim12.5$ at $z=4$. Thus at high redshift, the peak efficiency
of converting baryons into stars occurs in more massive haloes. The normalization of the SHM ratio decreases with increasing
redshift: While at $z=0$ the most efficient haloes have converted $\sim23$ per cent of their baryons into stars,
at $z=4$ the peak efficiency is less than $\sim10$ per cent.

We further compare our results to those of hydrodynamical zoom-in simulations and semi-analytic models and
find that most of the newer models show good agreement with the efficiencies needed to reproduce the
local observed SMFs. However, almost all have efficiencies which are far too high at earlier epochs, indicating
that they form their stars too early. At late times star formation is quenched, so that most of the stellar
mass today is already present at high redshift. In the semi-analytic models the normalization and the
slopes of the SHM relation are in good agreement with ours, but the halo mass where the efficiency peaks moves to
lower rather than to higher masses with increasing redshift.

We then combine our relation with merger trees extracted from the simulations in order to predict the mean assembly
histories of the stellar mass components within dark matter haloes (in situ and ex situ formation). We average
over a sample of haloes with the same mass at $z=0$, such that our model yields the average stellar mass growth of
central galaxies as a function of halo mass. We find that in low-mass systems the halo grows much more slowly than its central
galaxy. For a system with $\Mvir=10^{12}\Msun$, half of the stellar mass of the central galaxy forms
after redshift $z=0.7$ while half of the virial mass was already assembled by $z=1.3$. For massive systems the trends
are opposite at low redshift: the fractional growth of haloes is much stronger than stellar growth, while a large fraction of the
stellar mass of massive galaxies is already in place early. For a system with $\Mvir=10^{12}\Msun$, half of the halo has
assembled only by $z=1.0$, but half of the stellar mass was already in place by $z=1.8$. In very massive
haloes with $\Mvir=10^{12}\Msun$ the galaxies start to form very early with 25 per cent of their stellar mass in place already
by $z=3.0$. However, it takes them a long time to fully assemble: Half of their stellar mass is in place only by
$z=1.6$, while 75 per cent of the stellar mass is in place by $z=0.3$.

We then go further and identify the progenitors of a given halo at earlier epochs and derive the amount of stellar mass that has been
accreted onto the central galaxy. Employing a recipe to account for mass loss from individual stars, we then compute the stellar mass that has been
formed through star formation and convert this into a SFR. In this way we derive the SFR history as a function of halo mass.
We find that the star formation history of central galaxies peaks at a certain redshift which depends on halo mass: for massive systems the
peak occurs at high redshift while for low-mass systems it has not yet been reached by $z=0$. After the peak has been reached,
the SFR quickly drops, such that massive galaxies are quenched at $z\sim2$. On the other hand, the stellar accretion rates increase with time
for all systems. Massive systems have a high stellar accretion rate that dominates the total growth rate at late times, while for low-mass systems,
accretion is insignificant at all times. The halo mass at which the accretion rate and the SFR are equal at $z=0$ is  $\log(\Mvir/\Msun)\sim12.7$.

Finally we demonstrate that our inferred SFR histories are in agreement with observation. We compute the SFR for all systems in
the simulation and derive the evolution of the cosmic SFR density. Moreover we compute the specific SFR history for two different stellar mass
ranges. We find that our choice for the parameterization of the redshift-dependent SHM relation is consistent with observational constraints,
and that our inferred recipe for the mass dependent star formation histories gives reliable estimates.

\section*{Acknowledgements} 

We thank
Alyson Brooks,
Shy Genel,
Fabio Governato,
Qi Guo,
Cheng Li,
Takashi Okamoto,
Ludwig Oser,
Paola Santini,
Cecilia Scannapieco,
and Greg Stinson
for providing their data in electronic form.
The \textit{MS} and \textit{MS-II} were carried
out at the Computing Centre of the Max Planck Society in Garching.



\bibliographystyle{mn2e}
\bibliography{moster2011}

\begin{thebibliography}{}

\bibitem[\protect\citeauthoryear{{Abadi}, {Navarro} \& {Steinmetz}}{{Abadi}
  et~al.}{2006}]{abadi2006}
{Abadi} M.~G.,  {Navarro} J.~F.,    {Steinmetz} M.,  2006, \mnras, 365, 747

\bibitem[\protect\citeauthoryear{{Abazajian} et~al.,}{{Abazajian}
  et~al.}{2009}]{abazajian2009}
{Abazajian} K.~N.,  et~al., 2009, \apjs, 182, 543

\bibitem[\protect\citeauthoryear{{Angulo}, {Springel}, {White}, {Jenkins},
  {Baugh} \& {Frenk}}{{Angulo} et~al.}{2012}]{angulo2012}
{Angulo} R.~E.,  {Springel} V.,  {White} S.~D.~M.,  {Jenkins} A.,  {Baugh}
  C.~M.,    {Frenk} C.~S.,  2012, ArXiv e-prints

\bibitem[\protect\citeauthoryear{{Angulo} \& {White}}{{Angulo} \&
  {White}}{2010}]{angulo2010}
{Angulo} R.~E.,  {White} S.~D.~M.,  2010, \mnras, 405, 143

\bibitem[\protect\citeauthoryear{{Baldry}, {Glazebrook} \& {Driver}}{{Baldry}
  et~al.}{2008}]{baldry2008}
{Baldry} I.~K.,  {Glazebrook} K.,    {Driver} S.~P.,  2008, \mnras, 388, 945

\bibitem[\protect\citeauthoryear{{Behroozi}, {Conroy} \& {Wechsler}}{{Behroozi}
  et~al.}{2010}]{behroozi2010}
{Behroozi} P.~S.,  {Conroy} C.,    {Wechsler} R.~H.,  2010, \apj, 717, 379

\bibitem[\protect\citeauthoryear{{Bell}, {McIntosh}, {Katz} \&
  {Weinberg}}{{Bell} et~al.}{2003}]{bell2003}
{Bell} E.~F.,  {McIntosh} D.~H.,  {Katz} N.,    {Weinberg} M.~D.,  2003, \apjs,
  149, 289

\bibitem[\protect\citeauthoryear{{Benson}, {Lacey}, {Baugh}, {Cole} \&
  {Frenk}}{{Benson} et~al.}{2002}]{benson2002}
{Benson} A.~J.,  {Lacey} C.~G.,  {Baugh} C.~M.,  {Cole} S.,    {Frenk} C.~S.,
  2002, \mnras, 333, 156

\bibitem[\protect\citeauthoryear{{Berlind} \& {Weinberg}}{{Berlind} \&
  {Weinberg}}{2002}]{berlind2002}
{Berlind} A.~A.,  {Weinberg} D.~H.,  2002, \apj, 575, 587

\bibitem[\protect\citeauthoryear{{Binney} \& {Tremaine}}{{Binney} \&
  {Tremaine}}{1987}]{binney1987}
{Binney} J.,  {Tremaine} S.,  1987, {Galactic dynamics}.
Princeton, NJ, Princeton University Press, 1987, 747 p.

\bibitem[\protect\citeauthoryear{{Bouwens} et~al.,}{{Bouwens}
  et~al.}{2009}]{bouwens2009}
{Bouwens} R.~J.,  et~al., 2009, \apj, 705, 936

\bibitem[\protect\citeauthoryear{{Bouwens} et~al.,}{{Bouwens}
  et~al.}{2011}]{bouwens2011}
{Bouwens} R.~J.,  et~al., 2011, ArXiv e-prints

\bibitem[\protect\citeauthoryear{{Bower}, {Benson}, {Malbon}, {Helly}, {Frenk},
  {Baugh}, {Cole} \& {Lacey}}{{Bower} et~al.}{2006}]{bower2006}
{Bower} R.~G.,  {Benson} A.~J.,  {Malbon} R.,  {Helly} J.~C.,  {Frenk} C.~S.,
  {Baugh} C.~M.,  {Cole} S.,    {Lacey} C.~G.,  2006, \mnras, 370, 645

\bibitem[\protect\citeauthoryear{{Boylan-Kolchin}, {Ma} \&
  {Quataert}}{{Boylan-Kolchin} et~al.}{2008}]{boylan2008}
{Boylan-Kolchin} M.,  {Ma} C.-P.,    {Quataert} E.,  2008, \mnras, 383, 93

\bibitem[\protect\citeauthoryear{{Boylan-Kolchin}, {Springel}, {White},
  {Jenkins} \& {Lemson}}{{Boylan-Kolchin} et~al.}{2009}]{boylan2009}
{Boylan-Kolchin} M.,  {Springel} V.,  {White} S.~D.~M.,  {Jenkins} A.,
  {Lemson} G.,  2009, \mnras, 398, 1150

\bibitem[\protect\citeauthoryear{{Brooks} et~al.,}{{Brooks}
  et~al.}{2011}]{brooks2011}
{Brooks} A.~M.,  et~al., 2011, \apj, 728, 51

\bibitem[\protect\citeauthoryear{{Brown} et~al.,}{{Brown}
  et~al.}{2008}]{brown2008}
{Brown} M.~J.~I.,  et~al., 2008, \apj, 682, 937

\bibitem[\protect\citeauthoryear{{Bruzual} \& {Charlot}}{{Bruzual} \&
  {Charlot}}{2003}]{bruzual2003}
{Bruzual} G.,  {Charlot} S.,  2003, \mnras, 344, 1000

\bibitem[\protect\citeauthoryear{{Castellano} et~al.,}{{Castellano}
  et~al.}{2010}]{castellano2010}
{Castellano} M.,  et~al., 2010, \aap, 511, A20

\bibitem[\protect\citeauthoryear{{Chabrier}}{{Chabrier}}{2003}]{chabrier2003}
{Chabrier} G.,  2003, \pasp, 115, 763

\bibitem[\protect\citeauthoryear{{Cole} et~al.,}{{Cole}
  et~al.}{2001}]{cole2001}
{Cole} S.,  et~al., 2001, \mnras, 326, 255

\bibitem[\protect\citeauthoryear{{Conroy} \& {Wechsler}}{{Conroy} \&
  {Wechsler}}{2009}]{conroy2009}
{Conroy} C.,  {Wechsler} R.~H.,  2009, \apj, 696, 620

\bibitem[\protect\citeauthoryear{{Conroy}, {Wechsler} \& {Kravtsov}}{{Conroy}
  et~al.}{2006}]{conroy2006}
{Conroy} C.,  {Wechsler} R.~H.,    {Kravtsov} A.~V.,  2006, \apj, 647, 201

\bibitem[\protect\citeauthoryear{{Conroy}, {Wechsler} \& {Kravtsov}}{{Conroy}
  et~al.}{2007}]{conroy2007}
{Conroy} C.,  {Wechsler} R.~H.,    {Kravtsov} A.~V.,  2007, \apj, 668, 826

\bibitem[\protect\citeauthoryear{{Crain} et~al.,}{{Crain}
  et~al.}{2009}]{crain2009}
{Crain} R.~A.,  et~al., 2009, \mnras, 399, 1773

\bibitem[\protect\citeauthoryear{{Croton}, {Springel}, {White}, {De Lucia},
  {Frenk}, {Gao}, {Jenkins}, {Kauffmann}, {Navarro} \& {Yoshida}}{{Croton}
  et~al.}{2006}]{croton2006}
{Croton} D.~J.,  {Springel} V.,  {White} S.~D.~M.,  {De Lucia} G.,  {Frenk}
  C.~S.,  {Gao} L.,  {Jenkins} A.,  {Kauffmann} G.,  {Navarro} J.~F.,
  {Yoshida} N.,  2006, \mnras, 365, 11

\bibitem[\protect\citeauthoryear{{Cucciati} et~al.,}{{Cucciati}
  et~al.}{2012}]{cucciati2012}
{Cucciati} O.,  et~al., 2012, \aap, 539, A31

\bibitem[\protect\citeauthoryear{{Daddi} et~al.,}{{Daddi}
  et~al.}{2007}]{daddi2007}
{Daddi} E.,  et~al., 2007, \apj, 670, 156

\bibitem[\protect\citeauthoryear{{De Lucia} \& {Blaizot}}{{De Lucia} \&
  {Blaizot}}{2007}]{delucia2007}
{De Lucia} G.,  {Blaizot} J.,  2007, \mnras, 375, 2

\bibitem[\protect\citeauthoryear{{De Lucia}, {Springel}, {White}, {Croton} \&
  {Kauffmann}}{{De Lucia} et~al.}{2006}]{delucia2006}
{De Lucia} G.,  {Springel} V.,  {White} S.~D.~M.,  {Croton} D.,    {Kauffmann}
  G.,  2006, \mnras, 366, 499

\bibitem[\protect\citeauthoryear{{Dekel} \& {Silk}}{{Dekel} \&
  {Silk}}{1986}]{dekel1986}
{Dekel} A.,  {Silk} J.,  1986, \apj, 303, 39

\bibitem[\protect\citeauthoryear{{Drory}, {Salvato}, {Gabasch}, {Bender},
  {Hopp}, {Feulner} \& {Pannella}}{{Drory} et~al.}{2005}]{drory2005}
{Drory} N.,  {Salvato} M.,  {Gabasch} A.,  {Bender} R.,  {Hopp} U.,  {Feulner}
  G.,    {Pannella} M.,  2005, \apjl, 619, L131

\bibitem[\protect\citeauthoryear{{Dunne} et~al.,}{{Dunne}
  et~al.}{2009}]{dunne2009}
{Dunne} L.,  et~al., 2009, \mnras, 394, 3

\bibitem[\protect\citeauthoryear{{Fall} \& {Efstathiou}}{{Fall} \&
  {Efstathiou}}{1980}]{fall1980}
{Fall} S.~M.,  {Efstathiou} G.,  1980, \mnras, 193, 189

\bibitem[\protect\citeauthoryear{{Fazio} et~al.,}{{Fazio}
  et~al.}{2004}]{fazio2004}
{Fazio} G.~G.,  et~al., 2004, \apjs, 154, 10

\bibitem[\protect\citeauthoryear{{Feldmann}, {Carollo}, {Mayer}, {Renzini},
  {Lake}, {Quinn}, {Stinson} \& {Yepes}}{{Feldmann}
  et~al.}{2010}]{feldmann2010}
{Feldmann} R.,  {Carollo} C.~M.,  {Mayer} L.,  {Renzini} A.,  {Lake} G.,
  {Quinn} T.,  {Stinson} G.~S.,    {Yepes} G.,  2010, \apj, 709, 218

\bibitem[\protect\citeauthoryear{{Feulner}, {Gabasch}, {Salvato}, {Drory},
  {Hopp} \& {Bender}}{{Feulner} et~al.}{2005}]{feulner2005}
{Feulner} G.,  {Gabasch} A.,  {Salvato} M.,  {Drory} N.,  {Hopp} U.,
  {Bender} R.,  2005, \apjl, 633, L9

\bibitem[\protect\citeauthoryear{{Font}, {McCarthy}, {Crain}, {Theuns},
  {Schaye}, {Wiersma} \& {Dalla Vecchia}}{{Font} et~al.}{2011}]{font2011}
{Font} A.~S.,  {McCarthy} I.~G.,  {Crain} R.~A.,  {Theuns} T.,  {Schaye} J.,
  {Wiersma} R.~P.~C.,    {Dalla Vecchia} C.,  2011, \mnras, 416, 2802

\bibitem[\protect\citeauthoryear{{Fontana} et~al.,}{{Fontana}
  et~al.}{2006}]{fontana2006}
{Fontana} A.,  et~al., 2006, \aap, 459, 745

\bibitem[\protect\citeauthoryear{{Fontanot}, {De Lucia}, {Monaco}, {Somerville}
  \& {Santini}}{{Fontanot} et~al.}{2009}]{fontanot2009}
{Fontanot} F.,  {De Lucia} G.,  {Monaco} P.,  {Somerville} R.~S.,    {Santini}
  P.,  2009, \mnras, 397, 1776

\bibitem[\protect\citeauthoryear{{Genel} et~al.,}{{Genel}
  et~al.}{2012}]{genel2012}
{Genel} S.,  et~al., 2012, \apj, 745, 11

\bibitem[\protect\citeauthoryear{{Gnedin}}{{Gnedin}}{2000}]{gnedin2000}
{Gnedin} N.~Y.,  2000, \apj, 542, 535

\bibitem[\protect\citeauthoryear{{Governato} et~al.,}{{Governato}
  et~al.}{2012}]{governato2012}
{Governato} F.,  et~al., 2012, \mnras, p.~2697

\bibitem[\protect\citeauthoryear{{Guo} et~al.,}{{Guo}  et~al.}{2011}]{guo2011}
{Guo} Q.,  et~al., 2011, \mnras, 413, 101

\bibitem[\protect\citeauthoryear{{Guo}, {White}, {Li} \&
  {Boylan-Kolchin}}{{Guo} et~al.}{2010}]{guo2010}
{Guo} Q.,  {White} S.,  {Li} C.,    {Boylan-Kolchin} M.,  2010, \mnras, 404,
  1111

\bibitem[\protect\citeauthoryear{{Guo} \& {White}}{{Guo} \&
  {White}}{2008}]{guo2008}
{Guo} Q.,  {White} S.~D.~M.,  2008, \mnras, 384, 2

\bibitem[\protect\citeauthoryear{{Hambrick}, {Ostriker}, {Johansson} \&
  {Naab}}{{Hambrick} et~al.}{2011}]{hambrick2011}
{Hambrick} D.~C.,  {Ostriker} J.~P.,  {Johansson} P.~H.,    {Naab} T.,  2011,
  \mnras, 413, 2421

\bibitem[\protect\citeauthoryear{{Hatton}, {Devriendt}, {Ninin}, {Bouchet},
  {Guiderdoni} \& {Vibert}}{{Hatton} et~al.}{2003}]{hatton2003}
{Hatton} S.,  {Devriendt} J.~E.~G.,  {Ninin} S.,  {Bouchet} F.~R.,
  {Guiderdoni} B.,    {Vibert} D.,  2003, \mnras, 343, 75

\bibitem[\protect\citeauthoryear{{Hirschmann}, {Naab}, {Somerville}, {Burkert}
  \& {Oser}}{{Hirschmann} et~al.}{2012}]{hirschmann2012}
{Hirschmann} M.,  {Naab} T.,  {Somerville} R.~S.,  {Burkert} A.,    {Oser} L.,
  2012, \mnras, 419, 3200

\bibitem[\protect\citeauthoryear{{Hopkins} \& {Beacom}}{{Hopkins} \&
  {Beacom}}{2006}]{hopkins2006}
{Hopkins} A.~M.,  {Beacom} J.~F.,  2006, \apj, 651, 142

\bibitem[\protect\citeauthoryear{{Hopkins}, {Hernquist}, {Cox}, {Keres} \&
  {Wuyts}}{{Hopkins} et~al.}{2009}]{hopkins2009}
{Hopkins} P.~F.,  {Hernquist} L.,  {Cox} T.~J.,  {Keres} D.,    {Wuyts} S.,
  2009, \apj, 691, 1424

\bibitem[\protect\citeauthoryear{{Johansson}, {Naab} \& {Ostriker}}{{Johansson}
  et~al.}{2009}]{johansson2009}
{Johansson} P.~H.,  {Naab} T.,    {Ostriker} J.~P.,  2009, \apjl, 697, L38

\bibitem[\protect\citeauthoryear{{Jubelgas}, {Springel}, {En{\ss}lin} \&
  {Pfrommer}}{{Jubelgas} et~al.}{2008}]{jubelgas2008}
{Jubelgas} M.,  {Springel} V.,  {En{\ss}lin} T.,    {Pfrommer} C.,  2008, \aap,
  481, 33

\bibitem[\protect\citeauthoryear{{Kajisawa}, {Ichikawa}, {Yamada}, {Uchimoto},
  {Yoshikawa}, {Akiyama} \& {Onodera}}{{Kajisawa} et~al.}{2010}]{kajisawa2010}
{Kajisawa} M.,  {Ichikawa} T.,  {Yamada} T.,  {Uchimoto} Y.~K.,  {Yoshikawa}
  T.,  {Akiyama} M.,    {Onodera} M.,  2010, \apj, 723, 129

\bibitem[\protect\citeauthoryear{{Kang}, {Jing}, {Mo} \& {B{\"o}rner}}{{Kang}
  et~al.}{2005}]{kang2005}
{Kang} X.,  {Jing} Y.~P.,  {Mo} H.~J.,    {B{\"o}rner} G.,  2005, \apj, 631, 21

\bibitem[\protect\citeauthoryear{{Karim} et~al.,}{{Karim}
  et~al.}{2011}]{karim2011}
{Karim} A.,  et~al., 2011, \apj, 730, 61

\bibitem[\protect\citeauthoryear{{Katz}, {Weinberg} \& {Hernquist}}{{Katz}
  et~al.}{1996}]{katz1996}
{Katz} N.,  {Weinberg} D.~H.,    {Hernquist} L.,  1996, \apjs, 105, 19

\bibitem[\protect\citeauthoryear{{Kauffmann}}{{Kauffmann}}{1996}]{kauffmann1996}
{Kauffmann} G.,  1996, \mnras, 281, 487

\bibitem[\protect\citeauthoryear{{Kauffmann}, {Colberg}, {Diaferio} \&
  {White}}{{Kauffmann} et~al.}{1999}]{kauffmann1999}
{Kauffmann} G.,  {Colberg} J.~M.,  {Diaferio} A.,    {White} S.~D.~M.,  1999,
  \mnras, 303, 188

\bibitem[\protect\citeauthoryear{{Kauffmann}, {Li}, {Fu}, {Saintonge},
  {Catinella}, {Tacconi}, {Kramer}, {Genzel}, {Moran} \&
  {Schiminovich}}{{Kauffmann} et~al.}{2012}]{kauffmann2012}
{Kauffmann} G.,  {Li} C.,  {Fu} J.,  {Saintonge} A.,  {Catinella} B.,
  {Tacconi} L.~J.,  {Kramer} C.,  {Genzel} R.,  {Moran} S.,    {Schiminovich}
  D.,  2012, \mnras, 422, 997

\bibitem[\protect\citeauthoryear{{Kere{\v s}}, {Katz}, {Weinberg} \&
  {Dav{\'e}}}{{Kere{\v s}} et~al.}{2005}]{keres2005}
{Kere{\v s}} D.,  {Katz} N.,  {Weinberg} D.~H.,    {Dav{\'e}} R.,  2005,
  \mnras, 363, 2

\bibitem[\protect\citeauthoryear{{Khochfar} \& {Ostriker}}{{Khochfar} \&
  {Ostriker}}{2008}]{khochfar2008}
{Khochfar} S.,  {Ostriker} J.~P.,  2008, \apj, 680, 54

\bibitem[\protect\citeauthoryear{{Khochfar} \& {Silk}}{{Khochfar} \&
  {Silk}}{2006}]{khochfar2006}
{Khochfar} S.,  {Silk} J.,  2006, \mnras, 370, 902

\bibitem[\protect\citeauthoryear{{Klypin}, {Trujillo-Gomez} \&
  {Primack}}{{Klypin} et~al.}{2011}]{klypin2011}
{Klypin} A.~A.,  {Trujillo-Gomez} S.,    {Primack} J.,  2011, \apj, 740, 102

\bibitem[\protect\citeauthoryear{{Komatsu} et~al.,}{{Komatsu}
  et~al.}{2011}]{komatsu2011}
{Komatsu} E.,  et~al., 2011, \apjs, 192, 18

\bibitem[\protect\citeauthoryear{{Labb{\'e}} et~al.,}{{Labb{\'e}}
  et~al.}{2010}]{labbe2010}
{Labb{\'e}} I.,  et~al., 2010, \apjl, 716, L103

\bibitem[\protect\citeauthoryear{{Laporte}, {White}, {Naab}, {Ruszkowski} \&
  {Springel}}{{Laporte} et~al.}{2012}]{laporte2012}
{Laporte} C.~F.~P.,  {White} S.~D.~M.,  {Naab} T.,  {Ruszkowski} M.,
  {Springel} V.,  2012, ArXiv e-prints

\bibitem[\protect\citeauthoryear{{Larson}}{{Larson}}{1974}]{larson1974}
{Larson} R.~B.,  1974, \mnras, 169, 229

\bibitem[\protect\citeauthoryear{{Leauthaud} et~al.,}{{Leauthaud}
  et~al.}{2012}]{leauthaud2012}
{Leauthaud} A.,  et~al., 2012, \apj, 746, 95

\bibitem[\protect\citeauthoryear{{Li} \& {White}}{{Li} \&
  {White}}{2009}]{li2009}
{Li} C.,  {White} S.~D.~M.,  2009, \mnras, 398, 2177

\bibitem[\protect\citeauthoryear{{Ly}, {Lee}, {Dale}, {Momcheva}, {Salim},
  {Staudaher}, {Moore} \& {Finn}}{{Ly} et~al.}{2011}]{ly2011}
{Ly} C.,  {Lee} J.~C.,  {Dale} D.~A.,  {Momcheva} I.,  {Salim} S.,  {Staudaher}
  S.,  {Moore} C.~A.,    {Finn} R.,  2011, \apj, 726, 109

\bibitem[\protect\citeauthoryear{{Martizzi}, {Teyssier}, {Moore} \&
  {Wentz}}{{Martizzi} et~al.}{2012}]{martizzi2012}
{Martizzi} D.,  {Teyssier} R.,  {Moore} B.,    {Wentz} T.,  2012, \mnras,
  p.~2773

\bibitem[\protect\citeauthoryear{{Matteucci} \& {Francois}}{{Matteucci} \&
  {Francois}}{1989}]{matteucci1989}
{Matteucci} F.,  {Francois} P.,  1989, \mnras, 239, 885

\bibitem[\protect\citeauthoryear{{McLure} et~al.,}{{McLure}
  et~al.}{2011}]{mclure2011}
{McLure} R.~J.,  et~al., 2011, \mnras, 418, 2074

\bibitem[\protect\citeauthoryear{{Mo}, {Mao} \& {White}}{{Mo}
  et~al.}{1998}]{mo1998}
{Mo} H.~J.,  {Mao} S.,    {White} S.~D.~M.,  1998, \mnras, 295, 319

\bibitem[\protect\citeauthoryear{{Moster}, {Maccio'}, {Somerville}, {Naab} \&
  {Cox}}{{Moster} et~al.}{2011}]{moster2012}
{Moster} B.~P.,  {Maccio'} A.~V.,  {Somerville} R.~S.,  {Naab} T.,    {Cox}
  T.~J.,  2011, ArXiv e-prints

\bibitem[\protect\citeauthoryear{{Moster}, {Somerville}, {Maulbetsch}, {van den
  Bosch}, {Macci{\`o}}, {Naab} \& {Oser}}{{Moster} et~al.}{2010}]{moster2010}
{Moster} B.~P.,  {Somerville} R.~S.,  {Maulbetsch} C.,  {van den Bosch} F.~C.,
  {Macci{\`o}} A.~V.,  {Naab} T.,    {Oser} L.,  2010, \apj, 710, 903

\bibitem[\protect\citeauthoryear{{Moster}, {Somerville}, {Newman} \&
  {Rix}}{{Moster} et~al.}{2011}]{moster2011}
{Moster} B.~P.,  {Somerville} R.~S.,  {Newman} J.~A.,    {Rix} H.-W.,  2011,
  \apj, 731, 113

\bibitem[\protect\citeauthoryear{{Naab} \& {Burkert}}{{Naab} \&
  {Burkert}}{2003}]{naab2003}
{Naab} T.,  {Burkert} A.,  2003, \apj, 597, 893

\bibitem[\protect\citeauthoryear{{Naab}, {Johansson} \& {Ostriker}}{{Naab}
  et~al.}{2009}]{naab2009}
{Naab} T.,  {Johansson} P.~H.,    {Ostriker} J.~P.,  2009, \apjl, 699, L178

\bibitem[\protect\citeauthoryear{{Naab} \& {Ostriker}}{{Naab} \&
  {Ostriker}}{2006}]{naab2006}
{Naab} T.,  {Ostriker} J.~P.,  2006, \mnras, 366, 899

\bibitem[\protect\citeauthoryear{{Navarro} \& {Steinmetz}}{{Navarro} \&
  {Steinmetz}}{1997}]{navarro1997}
{Navarro} J.~F.,  {Steinmetz} M.,  1997, \apj, 478, 13

\bibitem[\protect\citeauthoryear{{Neistein}, {van den Bosch} \&
  {Dekel}}{{Neistein} et~al.}{2006}]{neistein2006}
{Neistein} E.,  {van den Bosch} F.~C.,    {Dekel} A.,  2006, \mnras, 372, 933

\bibitem[\protect\citeauthoryear{{Nipoti}, {Londrillo} \& {Ciotti}}{{Nipoti}
  et~al.}{2003}]{nipoti2003}
{Nipoti} C.,  {Londrillo} P.,    {Ciotti} L.,  2003, \mnras, 342, 501

\bibitem[\protect\citeauthoryear{{Noeske} et~al.,}{{Noeske}
  et~al.}{2007}]{noeske2007}
{Noeske} K.~G.,  et~al., 2007, \apjl, 660, L47

\bibitem[\protect\citeauthoryear{{Okamoto}}{{Okamoto}}{2012}]{okamoto2012}
{Okamoto} T.,  2012, ArXiv e-prints

\bibitem[\protect\citeauthoryear{{Oppenheimer} \& {Dav{\'e}}}{{Oppenheimer} \&
  {Dav{\'e}}}{2008}]{oppenheimer2008}
{Oppenheimer} B.~D.,  {Dav{\'e}} R.,  2008, \mnras, 387, 577

\bibitem[\protect\citeauthoryear{{Oppenheimer}, {Dav{\'e}}, {Kere{\v s}},
  {Fardal}, {Katz}, {Kollmeier} \& {Weinberg}}{{Oppenheimer}
  et~al.}{2010}]{oppenheimer2010}
{Oppenheimer} B.~D.,  {Dav{\'e}} R.,  {Kere{\v s}} D.,  {Fardal} M.,  {Katz}
  N.,  {Kollmeier} J.~A.,    {Weinberg} D.~H.,  2010, \mnras, 406, 2325

\bibitem[\protect\citeauthoryear{{Oser}, {Naab}, {Ostriker} \&
  {Johansson}}{{Oser} et~al.}{2012}]{oser2012}
{Oser} L.,  {Naab} T.,  {Ostriker} J.~P.,    {Johansson} P.~H.,  2012, \apj,
  744, 63

\bibitem[\protect\citeauthoryear{{Oser}, {Ostriker}, {Naab}, {Johansson} \&
  {Burkert}}{{Oser} et~al.}{2010}]{oser2010}
{Oser} L.,  {Ostriker} J.~P.,  {Naab} T.,  {Johansson} P.~H.,    {Burkert} A.,
  2010, \apj, 725, 2312

\bibitem[\protect\citeauthoryear{{Panter}, {Jimenez}, {Heavens} \&
  {Charlot}}{{Panter} et~al.}{2007}]{panter2007}
{Panter} B.,  {Jimenez} R.,  {Heavens} A.~F.,    {Charlot} S.,  2007, \mnras,
  378, 1550

\bibitem[\protect\citeauthoryear{{Peacock} \& {Smith}}{{Peacock} \&
  {Smith}}{2000}]{peacock2000}
{Peacock} J.~A.,  {Smith} R.~E.,  2000, \mnras, 318, 1144

\bibitem[\protect\citeauthoryear{{P{\'e}rez-Gonz{\'a}lez}
  et~al.,}{{P{\'e}rez-Gonz{\'a}lez}  et~al.}{2008}]{perez2008}
{P{\'e}rez-Gonz{\'a}lez} P.~G.,  et~al., 2008, \apj, 675, 234

\bibitem[\protect\citeauthoryear{{Prantzos} \& {Silk}}{{Prantzos} \&
  {Silk}}{1998}]{prantzos1998}
{Prantzos} N.,  {Silk} J.,  1998, \apj, 507, 229

\bibitem[\protect\citeauthoryear{{Puchwein} \& {Springel}}{{Puchwein} \&
  {Springel}}{2012}]{puchwein2012}
{Puchwein} E.,  {Springel} V.,  2012, ArXiv e-prints

\bibitem[\protect\citeauthoryear{{Robotham} \& {Driver}}{{Robotham} \&
  {Driver}}{2011}]{robotham2011}
{Robotham} A.~S.~G.,  {Driver} S.~P.,  2011, \mnras, 413, 2570

\bibitem[\protect\citeauthoryear{{Rujopakarn} et~al.,}{{Rujopakarn}
  et~al.}{2010}]{rujopakarn2010}
{Rujopakarn} W.,  et~al., 2010, \apj, 718, 1171

\bibitem[\protect\citeauthoryear{{Sales}, {Navarro}, {Theuns}, {Schaye},
  {White}, {Frenk}, {Crain} \& {Dalla Vecchia}}{{Sales}
  et~al.}{2011}]{sales2012}
{Sales} L.~V.,  {Navarro} J.~F.,  {Theuns} T.,  {Schaye} J.,  {White} S.~D.~M.,
   {Frenk} C.~S.,  {Crain} R.~A.,    {Dalla Vecchia} C.,  2011, ArXiv e-prints

\bibitem[\protect\citeauthoryear{{Salim} et~al.,}{{Salim}
  et~al.}{2007}]{salim2007}
{Salim} S.,  et~al., 2007, \apjs, 173, 267

\bibitem[\protect\citeauthoryear{{Santini} et~al.,}{{Santini}
  et~al.}{2012}]{santini2012}
{Santini} P.,  et~al., 2012, \aap, 538, A33

\bibitem[\protect\citeauthoryear{{Scannapieco}, {White}, {Springel} \&
  {Tissera}}{{Scannapieco} et~al.}{2011}]{scannapieco2011}
{Scannapieco} C.,  {White} S.~D.~M.,  {Springel} V.,    {Tissera} P.~B.,  2011,
  \mnras, 417, 154

\bibitem[\protect\citeauthoryear{{Schaerer} \& {de Barros}}{{Schaerer} \& {de
  Barros}}{2010}]{schaerer2010}
{Schaerer} D.,  {de Barros} S.,  2010, \aap, 515, A73

\bibitem[\protect\citeauthoryear{{Schaye} et~al.,}{{Schaye}
  et~al.}{2010}]{schaye2010}
{Schaye} J.,  et~al., 2010, \mnras, 402, 1536

\bibitem[\protect\citeauthoryear{{Shankar}, {Lapi}, {Salucci}, {De Zotti} \&
  {Danese}}{{Shankar} et~al.}{2006}]{shankar2006}
{Shankar} F.,  {Lapi} A.,  {Salucci} P.,  {De Zotti} G.,    {Danese} L.,  2006,
  \apj, 643, 14

\bibitem[\protect\citeauthoryear{{Smol{\v c}i{\'c}} et~al.,}{{Smol{\v c}i{\'c}}
   et~al.}{2009}]{smolcic2009}
{Smol{\v c}i{\'c}} V.,  et~al., 2009, \apj, 690, 610

\bibitem[\protect\citeauthoryear{{Somerville}, {Hopkins}, {Cox}, {Robertson} \&
  {Hernquist}}{{Somerville} et~al.}{2008}]{somerville2008}
{Somerville} R.~S.,  {Hopkins} P.~F.,  {Cox} T.~J.,  {Robertson} B.~E.,
  {Hernquist} L.,  2008, \mnras, 391, 481

\bibitem[\protect\citeauthoryear{{Spergel} et~al.,}{{Spergel}
  et~al.}{2003}]{spergel2003}
{Spergel} D.~N.,  et~al., 2003, \apjs, 148, 175

\bibitem[\protect\citeauthoryear{{Springel} et~al.,}{{Springel}
  et~al.}{2005}]{springel2005}
{Springel} V.,  et~al., 2005, \nat, 435, 629

\bibitem[\protect\citeauthoryear{{Springel} \& {Hernquist}}{{Springel} \&
  {Hernquist}}{2003}]{springel2003}
{Springel} V.,  {Hernquist} L.,  2003, \mnras, 339, 289

\bibitem[\protect\citeauthoryear{{Springel}, {White}, {Tormen} \&
  {Kauffmann}}{{Springel} et~al.}{2001}]{springel2001}
{Springel} V.,  {White} S.~D.~M.,  {Tormen} G.,    {Kauffmann} G.,  2001,
  \mnras, 328, 726

\bibitem[\protect\citeauthoryear{{Teyssier}, {Moore}, {Martizzi}, {Dubois} \&
  {Mayer}}{{Teyssier} et~al.}{2011}]{teyssier2011}
{Teyssier} R.,  {Moore} B.,  {Martizzi} D.,  {Dubois} Y.,    {Mayer} L.,  2011,
  \mnras, 414, 195

\bibitem[\protect\citeauthoryear{{Tinker}, {Weinberg}, {Zheng} \&
  {Zehavi}}{{Tinker} et~al.}{2005}]{tinker2005}
{Tinker} J.~L.,  {Weinberg} D.~H.,  {Zheng} Z.,    {Zehavi} I.,  2005, \apj,
  631, 41

\bibitem[\protect\citeauthoryear{{Tissera}, {White} \& {Scannapieco}}{{Tissera}
  et~al.}{2012}]{tissera2012}
{Tissera} P.~B.,  {White} S.~D.~M.,    {Scannapieco} C.,  2012, \mnras, 420,
  255

\bibitem[\protect\citeauthoryear{{Trenti} \& {Stiavelli}}{{Trenti} \&
  {Stiavelli}}{2008}]{trenti2008}
{Trenti} M.,  {Stiavelli} M.,  2008, \apj, 676, 767

\bibitem[\protect\citeauthoryear{{Vale} \& {Ostriker}}{{Vale} \&
  {Ostriker}}{2004}]{vale2004}
{Vale} A.,  {Ostriker} J.~P.,  2004, \mnras, 353, 189

\bibitem[\protect\citeauthoryear{{Vale} \& {Ostriker}}{{Vale} \&
  {Ostriker}}{2006}]{vale2006}
{Vale} A.,  {Ostriker} J.~P.,  2006, \mnras, 371, 1173

\bibitem[\protect\citeauthoryear{{van den Bosch}, {Yang} \& {Mo}}{{van den
  Bosch} et~al.}{2003}]{vdbosch2003}
{van den Bosch} F.~C.,  {Yang} X.,    {Mo} H.~J.,  2003, \mnras, 340, 771

\bibitem[\protect\citeauthoryear{{van der Burg}, {Hildebrandt} \& {Erben}}{{van
  der Burg} et~al.}{2010}]{vdburg2010}
{van der Burg} R.~F.~J.,  {Hildebrandt} H.,    {Erben} T.,  2010, \aap, 523,
  A74

\bibitem[\protect\citeauthoryear{{Wadepuhl} \& {Springel}}{{Wadepuhl} \&
  {Springel}}{2011}]{wadepuhl2011}
{Wadepuhl} M.,  {Springel} V.,  2011, \mnras, 410, 1975

\bibitem[\protect\citeauthoryear{{Wake} et~al.,}{{Wake}
  et~al.}{2011}]{wake2011}
{Wake} D.~A.,  et~al., 2011, \apj, 728, 46

\bibitem[\protect\citeauthoryear{{Werner} et~al.,}{{Werner}
  et~al.}{2004}]{werner2004}
{Werner} M.~W.,  et~al., 2004, \apjs, 154, 1

\bibitem[\protect\citeauthoryear{{White}}{{White}}{2001}]{white2001}
{White} M.,  2001, \mnras, 321, 1

\bibitem[\protect\citeauthoryear{{White}, {Zheng}, {Brown}, {Dey} \&
  {Jannuzi}}{{White} et~al.}{2007}]{white2007}
{White} M.,  {Zheng} Z.,  {Brown} M.~J.~I.,  {Dey} A.,    {Jannuzi} B.~T.,
  2007, \apjl, 655, L69

\bibitem[\protect\citeauthoryear{{White} \& {Rees}}{{White} \&
  {Rees}}{1978}]{white1978}
{White} S.~D.~M.,  {Rees} M.~J.,  1978, \mnras, 183, 341

\bibitem[\protect\citeauthoryear{{Windhorst} et~al.,}{{Windhorst}
  et~al.}{2011}]{windhorst2011}
{Windhorst} R.~A.,  et~al., 2011, \apjs, 193, 27

\bibitem[\protect\citeauthoryear{{Yang}, {Mo} \& {van den Bosch}}{{Yang}
  et~al.}{2003}]{yang2003}
{Yang} X.,  {Mo} H.~J.,    {van den Bosch} F.~C.,  2003, \mnras, 339, 1057

\bibitem[\protect\citeauthoryear{{Yang}, {Mo}, {van den Bosch}, {Zhang} \&
  {Han}}{{Yang} et~al.}{2011}]{yang2011}
{Yang} X.,  {Mo} H.~J.,  {van den Bosch} F.~C.,  {Zhang} Y.,    {Han} J.,
  2011, ArXiv e-prints

\bibitem[\protect\citeauthoryear{{York} et~al.,}{{York}
  et~al.}{2000}]{york2000}
{York} D.~G.,  et~al., 2000, \aj, 120, 1579

\bibitem[\protect\citeauthoryear{{Zehavi} et~al.,}{{Zehavi}
  et~al.}{2004}]{zehavi2004}
{Zehavi} I.,  et~al., 2004, \apj, 608, 16

\bibitem[\protect\citeauthoryear{{Zehavi} et~al.,}{{Zehavi}
  et~al.}{2011}]{zehavi2011}
{Zehavi} I.,  et~al., 2011, \apj, 736, 59

\bibitem[\protect\citeauthoryear{{Zheng}, {Bell}, {Papovich}, {Wolf},
  {Meisenheimer}, {Rix}, {Rieke} \& {Somerville}}{{Zheng}
  et~al.}{2007}]{zheng2007}
{Zheng} X.~Z.,  {Bell} E.~F.,  {Papovich} C.,  {Wolf} C.,  {Meisenheimer} K.,
  {Rix} H.-W.,  {Rieke} G.~H.,    {Somerville} R.,  2007, \apjl, 661, L41

\end{thebibliography}

\label{lastpage}

\end{document}